\documentclass[onecolum]{emulateapj}           

   \newcommand{\gsim}{\rlap{$>$}{\lower 1.0ex\hbox{$\sim$}}}

   \renewcommand{\sec}{\prime\prime}
   \renewcommand{\min}{\prime}

   \shorttitle{SFR and Optical Extinction in the FLS}
   \shortauthors{Choi et al}

   \normalsize

\begin{document}
\title{Star Formation Rates and Extinction Properties of IR-Luminous Galaxies in the {\it Spitzer} First Look Survey}

\author{P. I. Choi\altaffilmark{1}, L. Yan\altaffilmark{1}, 
  M. Im\altaffilmark{2}, G. Helou\altaffilmark{1}, 
  B. T. Soifer\altaffilmark{1}, L.J. Storrie-Lombardi\altaffilmark{1}, 
  R. Chary\altaffilmark{1}, H. I. Teplitz\altaffilmark{1},
  D. Fadda\altaffilmark{1}, 
  F. R. Marleau\altaffilmark{1}, M. Lacy\altaffilmark{1}, 
  G. Wilson\altaffilmark{1}, P. N. Appleton\altaffilmark{1}, 
  D. T. Frayer\altaffilmark{1}, and J. A. Surace\altaffilmark{1}}

\altaffiltext{1}{{\it Spitzer} Science Center, California Institute of
  Technology, MS~220-6, Pasadena, CA~91125; pchoi@\-ipac.caltech.edu,
  lyan@\-ipac.caltech.edu, gxh@\-ipac.caltech.edu, bts@\-ipac.caltech.edu,
  lisa@\-ipac.caltech.edu, rchary@\-caltech.edu, hit@\-ipac.caltech.edu, 
  fadda@\-ipac.caltech.edu, 
  marleau@\-ipac.caltech.edu, mlacy@\-ipac.caltech.edu,
  gillian@\-ipac.caltech.edu, apple@\-ipac.caltech.edu,
  frayer@\-ipac.caltech.edu, 
  jason@\-ipac.caltech.edu
}

\altaffiltext{2}{Seoul National University, Korea; mim@\-astroim.snu.ac.kr}

\begin {abstract}
  We investigate the instantaneous star formation rates ($SFR$) and
  extinction properties for a large ($N=274$), near-infrared
  (NIR: 2.2\micron) + mid-infrared (MIR: 24\micron) selected sample of
  normal to ultra-luminous infrared galaxies (ULIRGs) [$10^9 <
  L_{IR}/L_{\odot} < 10^{12.5}$] with $\langle z \rangle\sim0.8$ in
  the {\it Spitzer} Extragalactic First Look Survey (FLS).  We combine
  {\it Spitzer} MIPS 24\micron\, observations with high-resolution,
  optical Keck Deimos spectroscopy to derive optical emission-line
  ($H\alpha$,\,$H\beta$,\,[OII]) and infrared star formation rates
  ($SFR_{opt}$ \& $SFR_{IR}$, respectively).  Direct comparison of
  these SFR diagnostics reveals that our sample exhibits a wide range
  of extinction ($1.0<A_v<4.0$~mag).  This is after removing
  spectroscopic and IRAC color-selected AGN candidates that account
  for $\approx\!12\%$ of the sample. Objects with SFRs of a few solar
  masses per year have $A_v$ values consistent with those of normal
  spirals ($A_v\!\approx\!1.0$~mag).  By contrast, LIRGs at
  $z\gtrsim1$, which make up a large fraction of our sample, have
  $SFR\!\approx\!100$ $M_{\odot} yr^{-1}$ and a mean
  $A_v\!\approx\!2.5$~mag.  This translates to a 97\% mean attenuation
  of the [OII] $\lambda\lambda3727$ forbidden line doublet, with the
  most extreme sources having as much as 99.7\% of their [OII] line
  flux extinguished by dust.  Based on a $SFR_{IR}/SFR_{opt}$
  diagnostic, we derive an IR-luminosity-dependent $A_v^{IR}$ function
  [$A_v^{IR}=0.75*log(L_{IR}/L_{\odot})-6.35$~mag] that we use to
  extinction correct our emission line luminosities.  Application of
  this correction results in a correlation between $SFR_{IR}$ and
  $SFR_{opt}$ that has a dispersion of 0.2~dex (Semi-Interquartile
  Range).  Investigation of the $A_v$ dependence on redshift reveals
  that for a fixed $L_{IR}$ there is no significant $A_v$ evolution.
  Comparisons to previous studies reveal that the mean
  attenuation of our sample is intermediate between that of local
  optical/UV- and radio-selected samples and has a marginally stronger
  $L_{IR}$ dependence.
\end{abstract}

\section{Introduction}
Numerous investigations over the past decade have been directed at
measuring the cosmic star formation history (SFH) of the Universe.
Observations over the full spectrum from radio to X-ray have been
exploited to trace star formation rates \citep[see][for reviews of the
various diagnostics]{Kennicutt_98, Condon_92, Ghosh_White_01}.  The
most commonly utilized have historically been $H\alpha$ and [OII]
emission line and UV continuum flux \citep[eg.][]{Madau+96,
  Tresse_Maddox_98, Hogg+98, Yan+99, Glazebrook+99,
  Adelberger_Steidel_00, Erb+03}.  This is largely due to their
respective accessibility via ground-based observing windows for local
and distant samples and the fact that they are relatively direct
tracers of massive star formation.  Unfortunately, optical and UV
diagnostics are highly sensitive to dust attenuation.  Various
approaches have been implemented to estimate this reddening.  The use
of Balmer line flux ratios is a direct, but observationally taxing
method that requires high resolution spectroscopy.  In the absence of
multiple well measured emission lines, color- or magnitude-dependent
optical extinction corrections \citep{Rigopoulou+00,Hippelein+03}, or
the UV slope-extinction relation [$\beta$-$A_{UV}$] derived for
starburst galaxies \citep[eg.][]{Calzetti+94, Adelberger_Steidel_00}
have also been adopted.  Though commonly applied to non-starburst
galaxies, the latter has been shown to break down for both more and
less luminous systems \citep{Goldader+02, Bell+02, Bell_02}.

By comparison, far-infrared (FIR) and radio star formation rate (SFR)
diagnostics have the advantage that they are unaffected by extinction.
Their shortcoming, however, is that for general populations they have
more complex relationships to the star formation than optical emission
line and UV indicators.  For instance, the typically adopted FIR SFR
calibration \citep{Kennicutt_98} is based on the assumption of
infinite optical depth and 100\% reprocessing of massive star UV
emission into IR flux.  This is a reasonable assumption for heavily
extinguished systems, but breaks down for galaxies with moderate
attenuation.  In spiral galaxies for instance, counteracting effects
of UV radiation leakage and heating from older evolved stellar
populations must also be considered \citep{Lonsdale-Persson_Helou_87}.
Finally, despite our relatively limited understanding of the decimeter
radio continuum, it has served as a powerful proxy for the IR flux due to
the tightness of radio-FIR correlation \citep{Helou+85}.
Surprisingly, this correlation extends to IR luminosities
($\approx\!0.01$ L*) at which neither the bolometric infrared
luminosity ($L_{IR}$) nor the radio continuum are reliably tracing
star formation \citep{Bell_03}.

Comparative studies of these different diagnostics exist for a range
of galaxy types and sample selections.  UV and $H\alpha$ measurements
have been found to be generally consistent for local samples of normal
galaxies \citep{Sullivan+01, Bell_Kennicutt_01, Buat+02, Sullivan+04}.
The scatter and, in some cases, the offset between the tracers are
primarily attributed to a combination of uncertainties in the
extinction correction, star/dust geometry and the star formation
timescale \citep{Helou_Bicay_93, Bell_03, Sullivan+04}.  The
inter-comparison of UV/optical to decimeter radio emission
\citep{Sullivan+01, Afonso+03} and far-infrared IRAS observations
\citep{Cram+98, Dopita+02, Kewley+02, Hopkins+03} for large local
samples confirms the importance of accurate UV/optical attenuation
corrections.  Recent ISO-based studies by
\citet{Rigopoulou+00},\citet{Cardiel+03} and \citet{Flores+04} probe
out to more distant redshifts ($z\sim1$) but for admittedly small
samples of 12, 7 and 16 sources, respectively.  These pioneering works
provide the deepest IR-based SFR probes of the distant, dustier and
more actively starforming universe.  Consequently, they tend to
include more extreme, IR luminous galaxies than are present in the
local samples.

Due to the absence of a unified multiwavelength picture of the SFH,
there is considerable debate about the star formation density at high
redshift.  Most controversy revolves around issues of sample selection
effects, SFR calibration uncertainties and dust attenuation
corrections.  It has become clear that no single tracer is applicable
for all galaxies.  UV \& optical diagnostics appear to be well suited
for low IR luminosity galaxies that do not require significant
extinction corrections; whereas attenuation-free IR and radio
diagnostics provide better estimates in dusty, actively starforming
systems.  In this work we study the star formation and extinction
properties of a large, distant, actively starforming population by
making a direct comparison of optical emission line and IR SFR
diagnostics.  Specifically, we use mid-IR observations from the {\it
Spitzer} Extragalactic First Look Survey and deep Keck optical
spectroscopic observations to achieve an order of magnitude increase
in sample size over previous high redshift ISO studies.

This paper is divided into the following sections.  A summary of the
various observational components and their basic analysis is given in
\S2.  Calculations of the optical and infrared star formation rates
are described in \S3.  Our approach for removing contaminating AGN is
outlined in \S4.  The comparison between the optical and IR SFRs and
our derived extinction corrections are discussed in \S5.  The main
points of this work are summarized in \S6.  Throughout the paper, we
adopt the cosmology of $\Omega_{m} = 0.27$, $\Omega_\Lambda = 0.73$
and $H_{\circ} = 70$~km s$^{-1}$ Mpc$^{-1}$.

\section{Observations and Reductions}
\label{obs}
The {\it Spitzer} Extra-galactic First Look Survey (FLS)\footnote{For
  details of the FLS observation plan and the data release, see
  http://ssc.spitzer.caltech.edu/fls.} region is a
$\approx\!4$~sq.~deg region centered around RA=$17^h18^m00^s$,
Dec=$59^{\circ}30^{\min}00^{\sec}$.  It is chosen to lie within the
continuous viewing zone (CVZ), have minimum cirrus and no bright radio
sources.  Observations of this field with each of the 4 IRAC and 3
MIPS imaging bands was one of the first science tasks undertaken by
{\it Spitzer}.  In addition to IR imaging, numerous ancillary datasets
including radio, optical and near-IR (NIR) data have been taken in
this field.  Brief descriptions of the datasets included in this study
are given below.

\subsection{Imaging}

\subsubsection{Optical}
Optical $R$-band imaging of the FLS was carried out using the MOSAIC-I
camera on the 4-meter Mayall Telescope at the Kitt Peak National
Observatory on four consecutive nights on UT 2000 May 4-7.  A
$4\times2$ array of SITe $2048\times4096$ CCDs provides a
$36^{\min}\times36^{\min}$ field of view with a $0.258^{\sec}$ pixel
scale.  Tiling 26 individual pointings we obtain an $R$-band coverage
of 9.4~sq.~deg, with median exposure time per pointing of 1800 sec
and typical seeing FWHM$\sim\!1.0^{\sec}$.  The resulting source
catalog has a 50\% completeness limit of $R=24.5$~mag (Vega).  In
addition, $g\arcmin$ \& $i\arcmin$-band observations of the FLS were
obtained using the Large Format Camera (LFC) on the 200-inch Palomar
Observatory Hale Telescope.  These observations were taken over
multiple observing campaigns from August 2001 through June 2004.  The
total area coverage in these bands is roughly 2.0~sq.~deg, with
comparable seeing and resolution to that of the $R$-band data.
Comprehensive descriptions of these datasets are presented in
~\cite{Fadda+04} and Glassman et al. (2005, in preparation).

\subsubsection{Near-infrared}
Near-infrared observations were carried out in two separate, but
complimentary observing campaigns that can be characterized as
shallow, wide-field and deep, narrow-field.  In the first, $K_s$-band
imaging of a 1.14~sq.~deg region, to a median depth of $K_s<19.0$~mag
(Vega) was performed on UT 2001 May 23-26 using the Florida
Multi-object Imaging Near-IR Grism Observational Spectrometer
(FLAMINGOS) on the Kitt Peak National Observatory 2.1-m telescope.  A
$2048\times2048$ HgCdTe Rockwell array provides a
$20^{\min}\times20^{\min}$ field of view with $0.6^{\sec}$ pixel
scale.  Each pointing was comprised of 50, 30-sec exposures taken in a
25-position dither pattern.  The field was mapped with a $5\times5$
grid pattern using half-field offsets ($10^{\min}$) between pointings.
The median exposure time is 2400~sec per pixel, and the median stellar
PSF over the mosaic is FWHM=$1.6^{\sec}$.

In addition to the KPNO dataset, a smaller
$\approx\!45^{\min}\times45^{\min}$ verification region in the center of the FLS
was observed with the Wide-field Infrared Camera (WIRC) on the Palomar
Observatory Hale 200-inch telescope.  Observations were undertaken
over the course of multiple observing runs between June 2002 and July
2004.  A $2048\times2048$ Hawaii-II HgCdTe array provides a
$8.7^{\min}\times8.7^{\min}$ field of view with a $0.25^{\sec}$ pixel
scale.  The $\approx\!0.6$~sq.~deg area centered on the FLS
verification region was covered with 34 tiled pointings.  The average
exposure time per pointing is 3600~sec ($120\times30$sec) taken with a
30-position random dither pattern, to a median depth of $K_s<20.2$~mag
(Vega).  A detailed description of all of the NIR observations and
reductions is presented in Glassman et al. (2005, in preparation).

\subsubsection{Mid-Infrared}
The extragalactic component of the {\it Spitzer} FLS is comprised of
Infrared Array Camera (IRAC) \citep{Fazio+04} and Multiband Imaging
Photometer for {\it Spitzer} (MIPS) \citep{Rieke+04} observations taken in
December 2003 with a total exposure time of 63 hours.  The MIPS
24\micron\, area coverage was 4.4~sq.~deg for the main field and
0.26~sq.~deg in a deeper verification field, with respective
3$\sigma$ depths of $0.11/0.08$~mJy.  All data were processed and
stacked by the data processing pipeline at the {\it Spitzer} Science
Center (SSC).  MIPS photometry was performed using StarFinder
\cite{Diolaiti+00}, which measures profile-fitted fluxes for point
sources.  A complete description of the 24\micron\, data reduction and
source catalog can be found in \cite{Marleau+04} and Fadda et al.
(2005, in preparation).

\subsection{Spectroscopy}
Optical spectroscopy was obtained with the Deep Imaging Multi-Object
Spectrograph \citep[DEIMOS;][]{Faber+03} on the W. M. Keck II 10-meter
telescope.  Observations were performed over 3 nights from UT 2003
June 27-29.  A 1200 line mm$^{-1}$ grating with central wavelength
settings of 7400\AA\, and 7699\AA\, was used with the GG495 blocking
filter, resulting in a 0.33\AA\, pix$^{-1}$ mean spectral dispersion
and a 1.45\AA\, instrumental resolution.  The total spectral range
observed was $6300-9300$\AA; however, the coverage for an individual
source was limited to 2,630\AA\, with a slit position dependent
starting wavelength.

\begin{figure}[!thp]
\epsscale{0.8}
\plotone{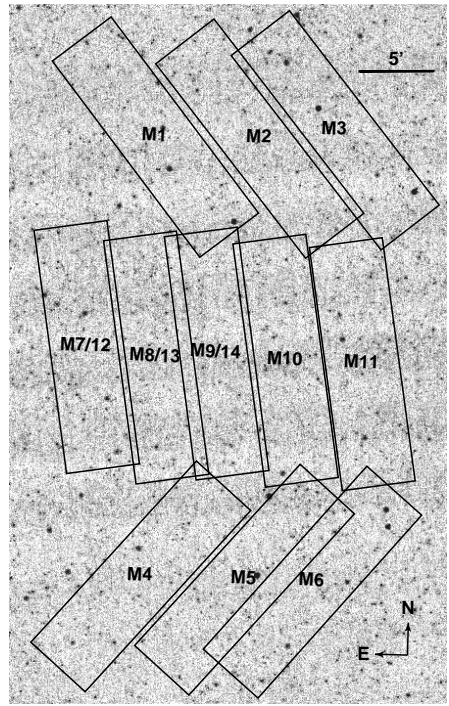}
\epsscale{1.0}
\caption{MIPS 24\micron\, image of the central
  $30^{\min}\times45^{\min}$ region of the First Look Survey
  Verification region.  Positions of 14 $5^{\min}\times17^{\min}$ Keck
  Deimos optical spectroscopy slitmasks are shown.  A summary of the
  spectroscopic observations is given in Table~1.
  \label{deimos_masks}}
\end{figure}

A total of 14 multi-slit masks were observed, with $\approx\!100$
$1^{\sec}$ wide slits per mask.  The $5^{\min}\times16^{\min}$
slitmasks were tiled in 11 unique positions to sample a
$25^{\min}\times45^{\min}$ area centered on the FLS.  In
Figure~\ref{deimos_masks}, mask positions are shown on top of the
24\micron\, mosaic for illustration.  Multiple masks were observed for
three of the positions in the deepest central region.  Table~1 lists
the positions, PAs and exposure times for the 14 observed masks.

\begin{deluxetable}{ccccc}
  \tablewidth{0pt} \tabletypesize{\scriptsize}
  \tablecaption{Spectroscopic Observation Log} \tablehead{ \colhead{ID} & \colhead{RA (J2000)} & \colhead{Dec (J2000)} & \colhead{P.A. (deg)} &
    \colhead{Exposure Time}} \startdata
 M1 &  17:17:08.51 &   59:59:53.5 &  -39 & $3\times20$~min \\
 M2 &  17:16:11.74 &   60:00:02.3 &  -40 & $3\times20$~min \\
 M3 &  17:15:30.86 &   60:00:46.0 &  -40 & $3\times20$~min \\
 M4 &  17:17:07.65 &   59:29:58.6 &   40 & $3\times20$~min \\
 M5 &  17:16:12.00 &   59:30:00.0 &   40 & $3\times20$~min \\
 M6 &  17:15:35.57 &   59:29:58.8 &   39 & $3\times20$~min \\
 M7 &  17:17:41.53 &   59:45:26.1 &  -10 & $5\times45$~min \\
 M8 &  17:17:03.68 &   59:45:00.0 &  -10 & $4\times45$~min \\
 M9 &  17:16:31.14 &   59:45:24.8 &  -10 & $4\times45$~min \\
 M10 &  17:15:53.81 &   59:45:05.2 &  -10 & $3\times20$~min \\
 M11 &  17:15:12.35 &   59:45:00.6 &  -10 & $3\times20$~min \\
 M12 &  17:17:41.39 &   59:45:32.0 &  -10 & $3\times20$~min \\
 M13 &  17:17:03.68 &   59:45:00.0 &  -10 & $3\times20$~min \\
 M14 &  17:16:31.14 &   59:45:24.8 &  -10 & $3\times20$~min \\
  \enddata 
\end{deluxetable}

We require minimum slit lengths of $7^{\sec}$ for local sky
subtraction and adopt the DEEP2 recommended strategy of using tilted
slits to better sample and remove sky lines.  Slit position angles
$\theta$ were required to be $10<|\theta|<25$ degrees from the spatial
axis and when possible, were positioned along the major axis of
elongated galaxies.  Of the 14~masks, 3 were observed for a total of
10,800~sec ($3\times3600$sec) and the remaining 10~masks were observed
for a total of 3,600~sec ($3\times1200$sec).  Observing conditions
over the course of the run were good, with typical seeing
FWHM$\sim\!0.7^{\sec}$.  For data reduction, we employ the DEEP2
spec2d pipeline\footnote{See
  http://astron.berkeley.edu/cooper/deep/spec2d/}, which is
based on the SDSS spectral reduction package.  This package performs
cosmic ray removal, flat-fielding, co-addition, sky subtraction,
wavelength calibration and both 2-d and 1-d spectral extraction.

\subsubsection{Emission-line measurements and Redshift Identification}
\label{emission_line_measurements}
The 1-d spectral output of the DEEP2 pipeline are analyzed using an
in-house IDL package written by PIC \& DF. Galaxy redshifts are
identified through visual inspection and galaxy template cross
correlation.  Line flux, equivalent width and kinematic measurements
are made by performing Gaussian profile fits to the 1-d spectra.  This
is done interactively with the user identifying the lines to be fit
and specifying the spectral region over which to perform the
line+continuum fit.  Our high spectral resolution (1.45\AA\,
instrumental resolution) allows for most lines to be modeled
individually.  One exception is the [OII] doublet $\lambda\lambda3726,
3729$, which is unresolved in $\approx\!1/2$ of our sample.  Rather
than measure each line independently, we adopt a double Gaussian
profile with a fixed line separation of [$(1+z)\times2.75$\AA].  We
require the two lines to have the same FWHM, but allow their flux
ratio to be a free parameter.

\begin{figure*}[!thp]
\epsscale{1.00}
\plotone{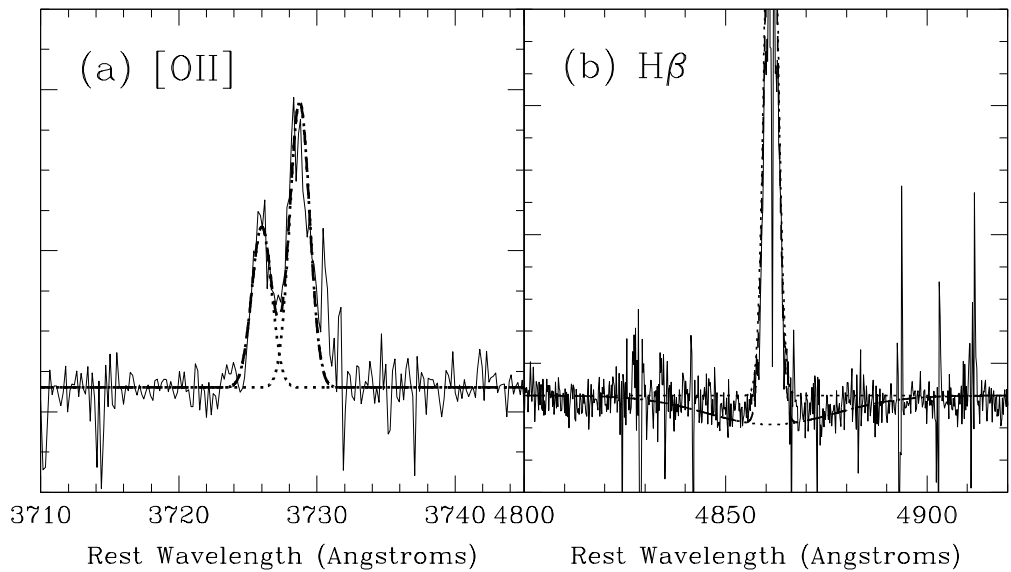}
\epsscale{1.0}
\caption{Line fits of [OII] and $H\beta$ for two representative
  objects in our sample.  In panel~({\it a}), the [OII] $\lambda\lambda3727$
  doublet is shown.  The gaussian profiles of the individual lines
  ({\it dotted}) are shown along with their co-addition ({\it
    dashed}).  The lines are required to have the same FWHM and a
  fixed 2.75\AA\, {\it restframe} separation.  In panel~({\it b}),
  individual components ({\it dotted}) and the coadded $H\beta$
  line-fit ({\it dashed}) are shown.
\label{OII_Hbeta}}
\end{figure*}

In the case of Balmer line fits, two-component emission+absorption
profiles are adopted.  It is well known that nebular Balmer emission
lines can suffer from contamination due to underlying stellar
absorption and subsequently be underestimated.  The standard technique
to correct for stellar absorption is to apply a global Balmer line
equivalent width correction.  Fortunately, our high spectral
resolution enables us to resolve and directly measure the nebular
emission and the pressure broadened stellar absorption components.
This eliminates the need for a global correction and results in more
accurate line fluxes.  Figure~\ref{OII_Hbeta} provides an illustration
of two typical emission line fits.  Figure~\ref{OII_Hbeta}{\it a} shows a
double Gaussian line fit to the [OII] $\lambda\lambda3727$ doublet, while
Figure~\ref{OII_Hbeta}{\it b} shows an emission+absorption two-component fit
to $H\beta$.

\subsubsection{Final MIR+NIR Sample}
The spectroscopic observations were designed to target a flux-limited
NIR-selected sample based primarily on our $K_s$-band dataset.
Optical $g\arcmin$-, $R$- \& $i\arcmin$-band color information was
included exclusively to a) clean out stellar contaminants and b) apply
a rough photometric redshift selection, designed to prioritize high
redshift sources with $z>0.6$.  The detailed description of this color
selection will be discussed separately in the spectroscopic catalog
paper.

\begin{figure}[!thp]
\epsscale{1.00}
\plotone{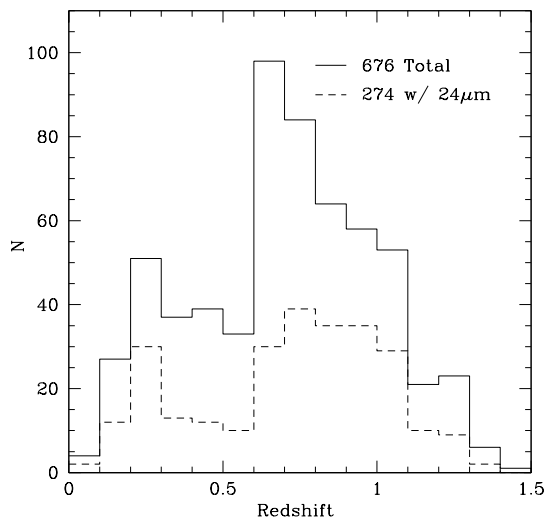}
\caption{Spectroscopic redshift distribution of our full 676 galaxy sample ({\it solid}) and the 274 galaxy, 24\micron-detected subsample ({\it dashed}).
\label{redshift_dist}}
\end{figure}

The final sample used for this study includes all galaxies in our
spectroscopic sample with a high confidence redshift and a significant
24\micron\, detection.  
In Figure~\ref{redshift_dist}, the full NIR-selected sample with good
spectroscopic redshifts ({\it solid}) is shown in contrast to the
subsample that have 24\micron-detected counterparts ({\it dashed}).

The effect of our $g\arcmin,R,i\arcmin,K_s$ color selection is seen in
both distributions as a sharp dropoff below $z<0.6$.  Beyond $z>1.1$
the impact of our decreasing spectroscopic redshift sensitivity is
evident.  In this regime strong sky lines make it challenging to
cleanly identify [OII] $\lambda\lambda3726,3729$.  By $z>1.3$, this
feature has moved out of our spectral coverage window for much of our
sample.  The final $K_s$+24\micron\, sample includes 274 galaxies and
has a median redshift of $\tilde{z}=0.76$.  

\begin{figure*}[!thp]
\epsscale{1.00}
\plotone{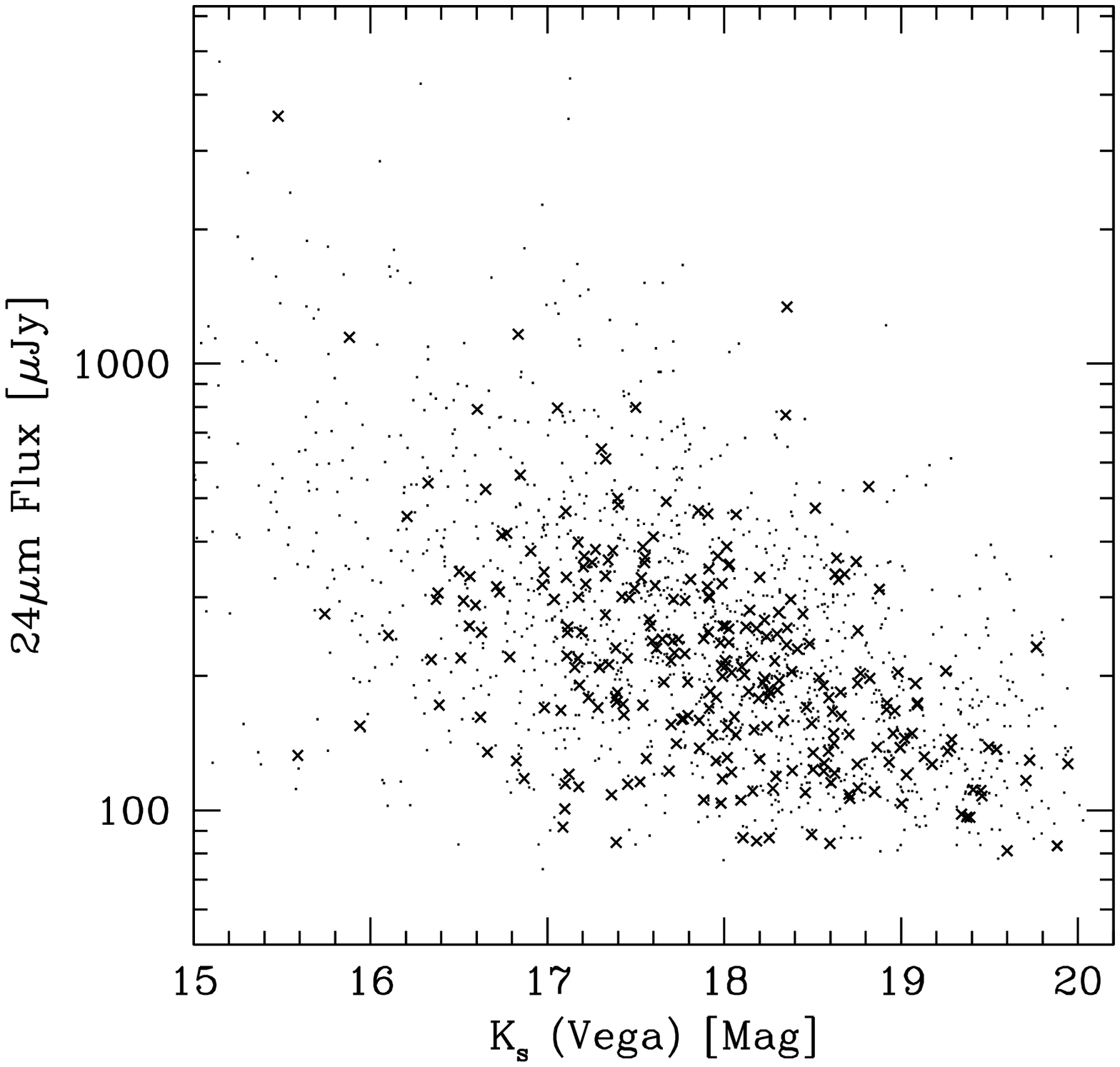}
\caption{MIPS 24\micron\, flux versus $K_s$-band mag for the 
  spectroscopic ({\it crosses}) and parent photometric ({\it
    dots}) samples.  Comparison of the two distributions illustrates
  that the parent population is being reasonably well represented
  with our spectroscopic sample for all but the brightest sources
  ($K_s>16.5$~mag).
\label{f24k}}
\end{figure*}

To investigate how our spectroscopic redshift target selection is
biasing our NIR+MIR distribution, we show in Figure~\ref{f24k}, the
$K_s$ versus 24\micron\, flux for the parent photometric sample ({\it
  dots}) and the targeted spectroscopic sample ({\it crosses}).  Our
selection against low redshift sources is evident in this distribution
by the lack of NIR bright ($K_s<16.5$~mag) spectroscopic targets.

\section{SFR Computations}
\label{section_sfr}

\subsection{Emission-Line SFR diagnostics}
\label{section_sfr_opt}
The $H\alpha$ $\lambda 6563$ emission line is the most prominent and
often measured SFR diagnostic in local galaxy surveys.  Unfortunately, 
at even moderate redshifts ($z\gtrsim0.4$), this line
becomes observationally taxing as it moves first into 
the forest of OH sky lines and
then beyond the optical spectral window.  Higher order Balmer lines
such as $H\beta$ and $H\gamma$ can be related to $H\alpha$;
however, these are often overlooked because their weak line strength
and contamination from underlying stellar absorption make them
difficult to measure.  At redshifts where all the prominent Balmer
lines become inaccessible ($z\gtrsim0.9$),
the [OII] doublet $\lambda\lambda3727$ can be used based on the $H\alpha$/[OII]
line ratio.  For this study, we use a combination of all these lines
to track the SFR over the redshift range of our sample.  $H\alpha$,
$H\beta$ and [OII] are measured for galaxies in the respective
redshift ranges $z\lesssim0.4$, $0.3\gtrsim z \gtrsim0.9$, and
$z\gtrsim 0.7$.  When accessible, $H\gamma$ and $H\delta$ are also
used as secondary diagnostics.  In this section, we discuss the flux
measurement and SFR conversion for each of these lines.

\subsubsection{Line Luminosity Measurement}
The most direct approach for measuring the total emission line flux of
a galaxy is direct integration of a flux-calibrated spectrum.
Unfortunately, this is impractical for many surveys, given the
challenges of properly flux calibrating large multi-slit samples.  An alternate
route is to combine line equivalent widths with broadband photometry
\citep*[eg.][]{Hogg+98,Hopkins+03}.  In our case, we compute $k$-corrected absolute
$u\arcmin$,$g\arcmin$ \& $R$-band magnitudes using our observed
$g\arcmin$,$R$, $i\arcmin$ and $K_s$ photometry and the KCORRECT code
\citep{Blanton+03}, adapted for our dataset.  We adopt these restframe
magnitudes as approximations of the continuum flux density at the rest
wavelengths of [OII], $H\beta$ and $H\alpha$ (3727\AA, 4861\AA\ and
6563\AA, respectively).  Luminosities of these emission lines are then
given by:

\begin{equation}
L_{H{\alpha}}=EW_{H{\alpha}}10^{[-0.4(M_{R}-34.10)]}\frac{3.0\times10^{18}}{\lambda_{rest}^{2}}~~~~(W)
\end{equation}
\begin{equation}
L_{H{\beta}}=EW_{H{\beta}}10^{[-0.4(M_{g'}-34.10)]}\frac{3.0\times10^{18}}{\lambda_{rest}^{2}}~~~~(W) 
\end{equation}
\begin{equation}
L_{[OII]}=EW_{[OII]}10^{[-0.4(M_{u'}-34.10)]}\frac{3.0\times10^{18}}{\lambda_{rest}^{2}}~~~~(W)
\end{equation}

\noindent where $EW$ is the line equivalent width in 
angstroms; $M$ is the $k$-corrected absolute AB magnitude appropriate
for the rest wavelength of the line being measured; and
$\lambda_{rest}$ is the central wavelength of the broadband filter,
also in angstroms.  These derivations assume that the continuum flux
at a given emission line wavelength is well approximated by the flux
density at the corresponding broadband effective wavelength.  An
additional color correction can be made to account for the wavelength
difference between the line and the filter effective wavelength;
however, we find this correction to be negligibly small (of order a
few percent), so it is excluded.

A major benefit of this approach is that it alleviates the need for
aperture/slit corrections that can plague direct flux
measurements.  
These corrections are small for compact galaxies,
but can be substantial for extended sources.  Consequently, aperture
loss tends to be redshift dependent and has the potential to
masquerade as evolution.  
By constrast, line equivalent widths and the scaled line fluxes
described above are fairly insensitive to slit loss.  This approach
does rely on the underlying assumption that the average
line-to-continuum ratio within the slit is the same as that outside
the slit; however, the same assumption is made when correcting for
slit loss.

\subsubsection{Balmer Line SFR diagnostics}
The Balmer line star formation rate diagnostics
rely on the $H\alpha$-$SFR$
calibration derived 
by \cite{Kennicutt_98}:
\begin{equation}
SFR_{H\alpha}=7.90\times 10^{-42} L_{H\alpha}E_{H\alpha}~~~~(ergs~s^{-1})
\end{equation}

\noindent where $L_{H\alpha}$ is the measured line luminosity
and $E_{H\alpha}$ is the extinction correction factor, measured at
the wavelength of $H\alpha$.  This correlation is based on
evolutionary synthesis models assuming solar abundance, a Salpeter IMF
with stellar mass limits of $0.1<M<100M_{\odot}$, and $T_e=10,000$~K
case-B recombination.  It also assumes that the escape fraction of
ionizing radiation from the observed galaxy is negligible, and
therefore that the nebular emission traces all of the massive star
formation.

The conversion for higher order Balmer lines are derived based on the
intrinsic case-B recombination line ratios.  As discussed in
\S\ref{emission_line_measurements}, these lines tend to be less
frequently adopted as SFR tracers due the difficulties of properly
correcting for stellar absorption; however, \cite{Kennicutt_92} has
shown that even with the adoption of a mean correction, for strong
emission line galaxies $H\beta$ can be a reliable SFR diagnostic.
Incorporating the expected $H\alpha/H\beta=2.87$ line ratio
\citep{Osterbrock_89}, we obtain the following relation for
$SFR_{H\beta}$:

\begin{equation}
SFR_{H\beta}=2.75\times 10^{-42} L_{H\beta}E_{H\beta}~~~~(ergs~s^{-1})
\end{equation}

\noindent where the extinction term $E_{H\beta}$ is measured at $H\beta$.
Similar relations for $H\gamma$ and  $H\delta$ are derived based on the 
$H\gamma/H\beta=0.466$ and $H\delta/H\beta=0.256$ line ratios.

\subsubsection{[OII] Forbidden Line SFR diagnostic}
For the most distant $z\gtrsim 0.7$ sources in our sample, we compute
the SFR based on the [OII] $\lambda\lambda3727$ forbidden line doublet,
adopting the \cite{Kewley+04} calibration:

\begin{equation}
SFR_{[OII]}=(0.66\pm0.17)\times 10^{-41} L_{[OII]}E_{[OII]}~~~~(ergs~s^{-1})
\label{sfr_o2_kewley}
\end{equation}

\noindent Following the notation above $L_{[OII]}$ is the measured
line luminosity and $E_{[OII]}$ is the extinction correction factor at
$\lambda3727$.  An important difference with respect to most previous
[OII] calibrations is that Equation~\ref{sfr_o2_kewley} does not
include any assumptions about the differential reddening
between $H\alpha$ and [OII] of the source.  By contrast, the standard
conversion of \cite{Kennicutt_98}
\citep*[cf.][]{Gallagher+89,Kennicutt_92}:

\begin{equation}
\label{sfr_o2_kenn}
SFR_{[OII]}=(1.4\pm0.4)\times 10^{-41} L_{[OII]}E_{H\alpha}~~~~(ergs~s^{-1})
\end{equation}

\noindent requires the calibration of a mean reddening between
$H\alpha$ $\lambda 6563$ and [OII] $\lambda\lambda3727$.  The
assumption of an average $H\alpha$-to-[OII] reddening is reasonable
for samples with a narrow range of intrinsic extinction; however, it would lead
to systematic, extinction-dependent errors for more general samples
with broad $E(B-V)$ distributions.  Also, since the current sample has 
a considerably brighter mean IR luminosity than that of 
the \cite{Kennicutt_92} or \cite{Gallagher+89} calibration datasets, 
using Equation~\ref{sfr_o2_kenn} would tend to underestimate both 
the total [OII] extinction and star formation rate.

It is worth noting that Equation~\ref{sfr_o2_kewley} ignores
metallicity effects on the [OII]-to-Balmer line luminosity ratio.
{Kewley+04} does characterize the abundance dependence of this ratio;
however, due to our lack of metallicity measurements, we use their
$SFR_{[OII]}$ calibration that adopts a mean abundance value of the
Nearby Field Galaxy Survey (NFGS).  Though a detailed investigation of
this issue is beyond the scope of this work, we can use their findings
to estimate the impact of metal abundance on our derived SFRs.  Over
the metallicity range $8.0<log(O/H)+12<9.0$ \citep[based on][$R_{23}$
calibration]{McGaugh_91} of the NFGS sample, \citet{Kewley+04} find
that the [OII]/$H\alpha$ ratio exhibits a metallicity dependent range
of $\pm0.2$~dex.  The adoption of a mean abundance introduces
$\lesssim0.08$~dex of scatter into their [OII]/$H\alpha$ correlation.
We expect a comparable impact on the scatter of our $SFR_{[OII]}$
diagnostic.  At that level, it does not have a significant effect on
our $SFR([OII])$ or extinction uncertainties.

\begin{figure*}[!thp]
  \epsscale{1.00}
  \plotone{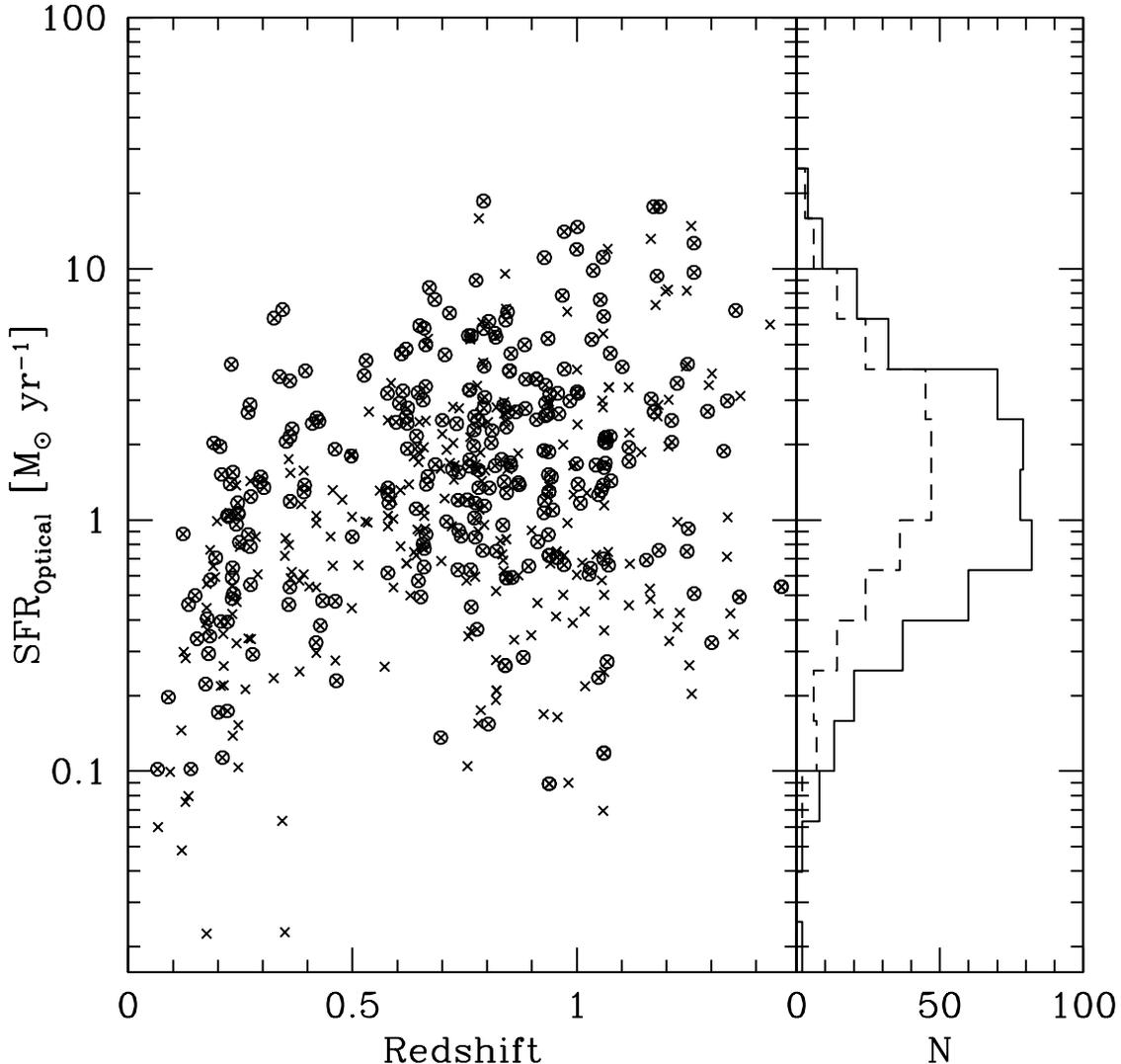}
  \epsscale{1.0}
  \caption{Extinction-uncorrected emission line-derived $SFR_{opt}$ versus redshift for the full 676 galaxy, NIR-selected sample ({\it crosses}) and the subset of 274 galaxy, $K_s$+24\micron-detected sample ({\it circles}).  The integrated $SFR_{opt}$ distributions of the full ({\it solid}) and 24\micron-detected ({\it dashed}) samples show that the mean $SFR_{opt}$ for IR-luminous sources is higher than that of the parent population.  The lack of a clean separation between the detected and non-detected sources is an expected consequence of a $L_{IR}$-dependent optical extinction.
\label{SFR_vs_z}}
\end{figure*}

Finally, in Figure~\ref{SFR_vs_z} we combine the above emission line
diagnostics to look at the optically-derived, extinction-uncorrected
SFR versus redshift.  Comparison of the parent NIR-selected sample
with the 24\micron-detected subsample reveals that latter has a higher
mean uncorrected $SFR_{opt}$.  This not an unexpected result, since IR
flux traces star formation.  A bit surprising is the degree
of overlap between the $SFR_{opt}$ distributions of IR-detected and
non-detected sources.  The fact that at any given redshift,
$SFR_{opt}$ (uncorrected) provides little indication of whether a
source will be IR luminous or not suggests an IR-luminosity
dependent attenuation.

\subsection{IR SFR Calculations}
\label{section_sfr_ir}
An alternative SFR tracer that is unaffected by extinction is the
infrared luminosity The IR component of galaxy SEDs can be decomposed
into three main dust emission components: 1) a near blackbody emission
profile of a thermally heated `cold' big grain (BG) dust component; 2)
a near blackbody component of a stochastically heated `warm' very
small grain (VSG) dust; and 3) molecular polycyclic aromatic
hydrocarbon (PAH) emission features \citep[][hereafter
DH02]{Desert+90, Dale_Helou_02}.  The primary heat sources for these
dust components are stellar radiation from young stars, an older
evolved stellar population and AGN.

In dusty, high-opacity systems where the dominant heat source of the
IR dust radiation is young stars (ie. Starbursts and LIRGs) the IR
luminosity is expected to be an excellent tracer of the instantaneous
SFR.  In these situations, the conversion of $L_{IR}$ to a star
formation rate can be made with the calibration of \cite{Kennicutt_98},

\begin{equation}
\label{sfr_ir}
SFR_{IR}=4.5\times 10^{-44} L_{IR} (ergs~s^{-1})
\end{equation}

\noindent where $L_{IR}$ is defined as the
integrated luminosity from $8-1000\micron$\footnote{nb. Different
  authors have conflicting definitions for $L_{IR}$ and $L_{FIR}$.}.
Equation~\ref{sfr_ir} is based on the assumption of solar abundance, a
Salpeter IMF ($0.1-100M_{\sun}$) and an optically thick dust
distribution.  It is consistent ($\pm 30\%$) with published
calibrations of comparably selected samples \citep{Kennicutt_98};
however, its extension to more general galaxy populations should be
made with caution.  Specifically, the assumption of high optical depth
places an important limitation on its application to normal spiral
galaxies.  Large UV/optical escape fractions, whether due to low dust
opacity or dust/star geometry, will cause $SFR_{IR}$ to underestimate
the true SFR.  On the other hand, heating of the diffuse ISM by an
older background population can contribute significantly to the IR
luminosity \citep{Lonsdale-Persson_Helou_87, Helou_00}, resulting in an
overestimate of the SFR.  At the high IR luminosity extreme, the
$SFR_{IR}$ calibration runs into the problem that many ULIRGs derive a
significant fraction of their bolometric luminosity from AGN.  In
these systems, dust heating from a central AGN can be the dominant
component of $L_{IR}$.  This contribution is difficult to quantify, so
it is essential to screen these sources.  In \S\ref{agn_sect} we
discuss our approach for removing AGN from our sample.

\subsubsection{Bolometric Correction}
\label{sect_bol_corr}
To calculate $SFR_{IR}$, we first compute the bolometric IR
luminosity from our observed 24\micron\, observations.  A standard
approach for deriving $L_{IR}$ for local galaxies is to use
the definition of \citet{Sanders_Mirabel_96}, based on IRAS 12, 25, 60
\& 100\micron\, luminosities:

\begin{eqnarray}
\lefteqn{L_{IR}[8-1000\mu m]=1.8\times10^{-14}\times10^{26}\times} \\
&&[13.48L_{\nu}(12)+ 5.16L_{\nu}(25) + 2.58L_{\nu}(60) +L_{\nu}(100)]\nonumber
\label{sanders_mirabel}
\end{eqnarray}

\noindent where $L_{\nu}(\lambda)$ is in units of $L_{\sun} Hz^{-1}$
and $L_{IR}$ is in units of $L_{\sun}$.  A comparable {\it Spitzer}
band transformation exists (DH02)\nocite{Dale_Helou_02};
however, its application for distant galaxies is hindered by the
general dearth of multi-band FIR photometry. For instance, the bulk of
our FLS sample is observed, but undetected with 70 \& 160\micron\,
imaging.  Fortunately, it has been shown from IRAS and ISOCAM data
that the MIR alone is a reasonable tracer of $L_{IR}$ \citep{Elbaz+02,
  Takeuchi+05}.  We exploit this finding and use our 24\micron\,
observations, which correspond to restframe $10-24\micron\,$ over the
redshift range of our sample, to derive IR luminosities.

\begin{figure*}[!thp]
  \epsscale{1.00}
  \plotone{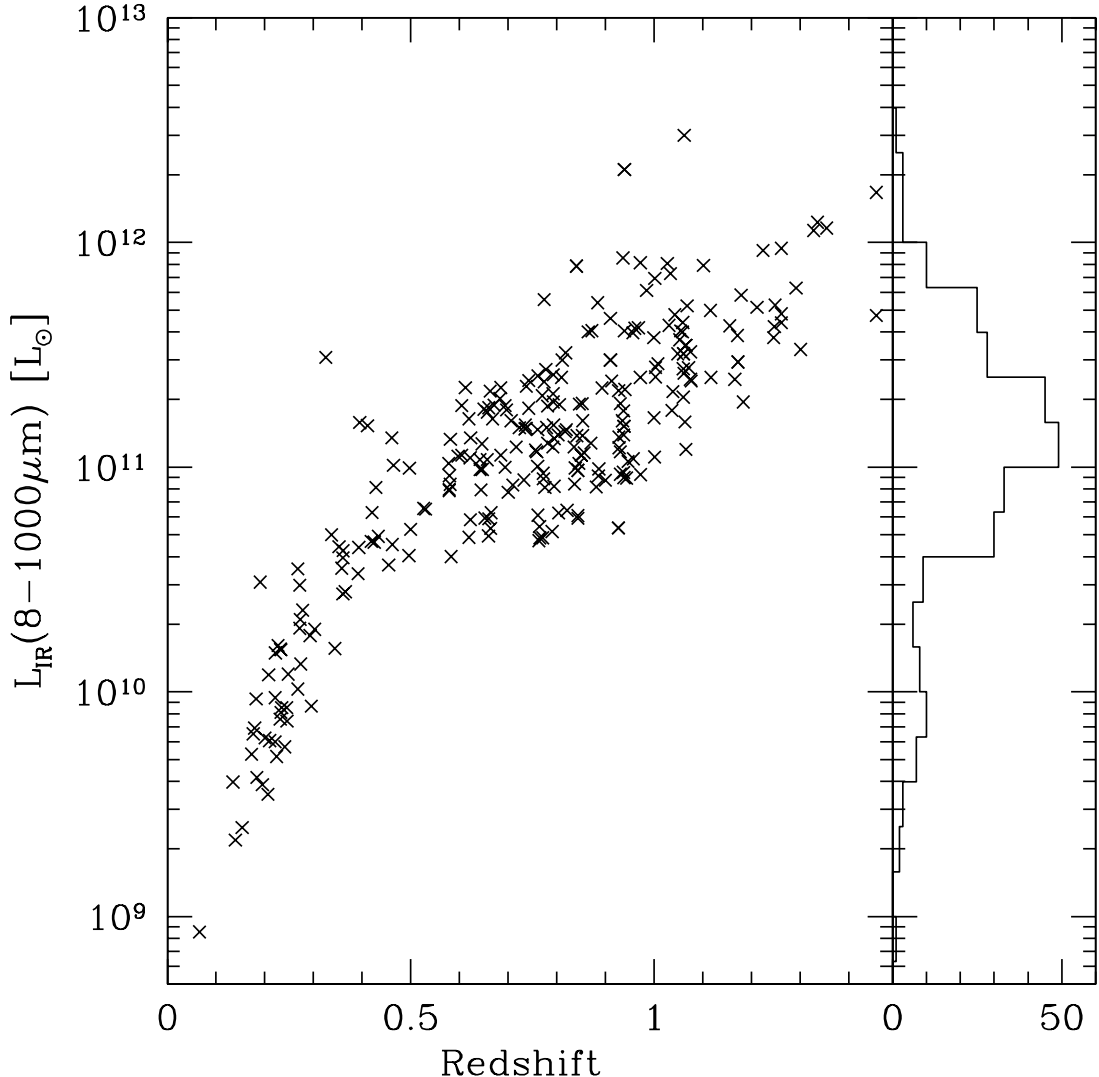}
  \epsscale{1.0}
\caption{$L_{IR}$ versus redshift for our full (N=274) $K_s$+24\micron-detected sample with accurate spectroscopic redshifts.  The $L_{IR}$ distribution is also shown in the {\it right} panel.
  \label{LIR_vs_z}}
\end{figure*}

Rather than implement a single $L_{MIR}$-to-$L_{IR}$ correlation, we
take the approach of using template SEDs \citep[][hereafter
CE01]{Chary_Elbaz_01} to derive $L_{IR}$[8--1000\micron] from the
24\micron\, fluxes.  CE01\nocite{Chary_Elbaz_01} have constructed
a library of 105 flux-calibrated SEDs that are single-valued in
$L_{IR}$ and cover the spectral range $0.1-1000$\micron.  They start
with model SEDs matched to a sample of local galaxies ranging from a
normal spiral (M51) to a ULIRG (Arp 220) \citep{Silva+98} and combine
them with MIR ISOCAM spectra and FIR model SEDs.  They then split and
recombine the MIR and FIR components of these templates to create a
library of SEDs that reproduce the observed correlations between the
various IRAS, ISOCAM and SCUBA IR fluxes of local galaxies.  We
implement these templates in a manner described by \cite{Elbaz+02} as
the `multi-template' approach.  For each galaxy in our sample, we
shift the template SEDs to the redshift of that source.  We choose the
template that most closely reproduces the observed 24\micron\, flux
and use its 12, 25, 60 and 100\micron\, integrated fluxes to compute
$L_{IR}$[8--1000\micron] based on Equation~\ref{sanders_mirabel}.  
The final distribution of $L_{IR}$ for the $K_s$+24\micron-detected
sample is shown in Figure~\ref{LIR_vs_z}.  This plot of $L_{IR}$
vs. redshift reveals that our sample spans a broad range in both
$L_{IR}$ and redshift but is strongly concentrated around
$L_{IR}/L_{\odot}\approx\!2.0\times10^{11}$ and $z\approx\!0.8$.
Also, as a consequence of our 24\micron\, flux limit, $L_{IR}$ and
redshift are correlated.  The impact of this correlation on our
results will be discussed further in \S\ref{redshift_dependence}.

It should noted that the application of this CE01 SED library relies
on the assumption that the luminosity trends seen locally are
representative of our sample, at higher redshift.  This has been shown
to be reasonable at least out to $z\sim0.8$ by \cite{Elbaz+02} based on a comparison of
their Radio-MIR vs. MIR-FIR correlations.  In
the next section we investigate the uncertainty in our
derived $L_{IR}$ using an independent family of model
SEDs.

\subsubsection{IR Bolometric Correction Uncertainty}
\label{sect_bol_corr_unc}
The absolute calibration and the intrinsic uncertainty of the mid-IR
to $L_{IR}$ correlation is investigated by comparing our computed IR
luminosities to those based on model SEDs of
DH02\nocite{Dale_Helou_02}.  Dale and collaborators generate
semi-empirical 3-1100\micron\, model SEDs for normal star-forming
galaxies.  These models are created by combining three-component
(large-grain, very small-grain and PAH) dust emission curves based on
a power-law distribution of the dust mass over heating intensity.  In
contrast to the CE01\nocite{Chary_Elbaz_01} family of SEDs, which are
based on a combination of slightly modified empirical SEDs, these are
built primarily from theoretical model emission curves in which the
PAH-dominated MIR (3-12\micron) region is replaced with a modified
ISOPHOT spectral component.  This family of SEDs is single-valued in
FIR color ($f_v(60\micron)$/$f_v(100\micron)$), indicating that a
galaxy IR SED can be uniquely determined with the measurement of this
single FIR flux ratio.

Since we lack FIR colors, rather than try to determine the best-fit
DH02\nocite{Dale_Helou_02} template for each galaxy, we compute
$L_{IR}$ for the entire family of SEDs, normalized to our observed
24\micron\, fluxes.  For this comparison, we adopt the conversion:

\begin{eqnarray}
\lefteqn{L_{IR}^{DH}[3-1100\micron]=[\zeta_1(z)\nu L_{\nu}(24\micron)} \\
&& \mbox{} + \zeta_2(z)\nu L_{\nu}(70\micron) + \zeta_3(z)\nu L_{\nu}(160\micron)]\nonumber
\end{eqnarray}

\noindent{where} the coefficients [$\zeta_1(z)$,$\zeta_2(z)$,$\zeta_3(z)$] 
are taken from DH02\nocite{Dale_Helou_02}.  These SEDs represent the
span of star-forming galaxy types, so the range of $L_{IR}^{DH}$
provides an estimate of the bolometric correction error due to our IR
SED assumptions.  

\begin{figure*}[!thp]
  \epsscale{1.00}
  \plotone{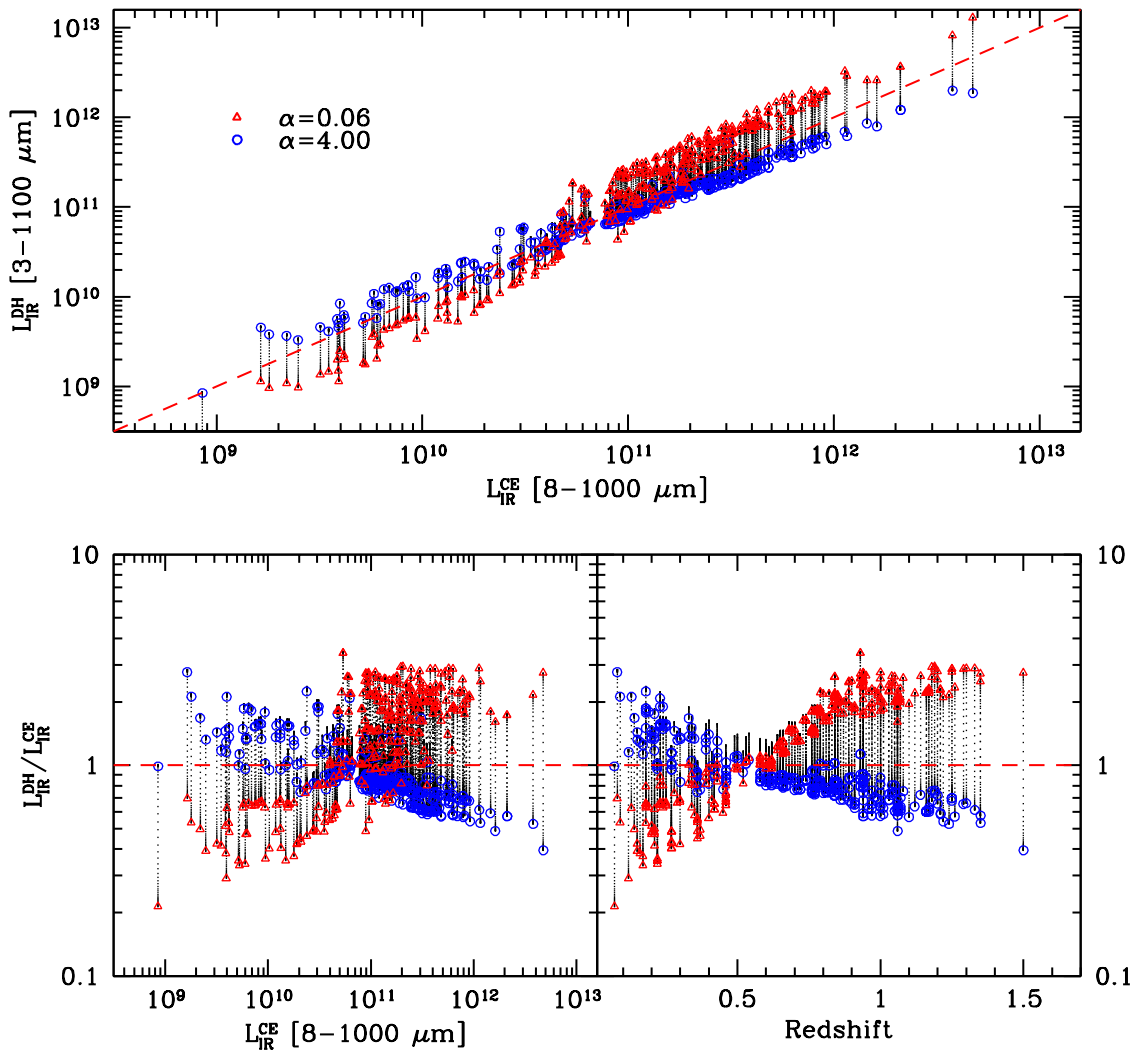}
  \epsscale{1.0}
\caption{Comparison of two independent derivations of the infrared 
  luminosity $L_{IR}$ for sources in our sample.  In the {\it top}
  panel, the best-fit $L_{IR}^{CE}$ for each source in our sample is
  compared to a range of $L_{IR}^{DH}$ computed for the full family of
  DH02 SEDs ({\it black dots}).  Open symbols
  represent the limiting `hot' ($\alpha=$0.06; {\it red triangles})
  and `cold' ($\alpha=$4.00; {\it blue circles}) DH02 SEDs.  The {\it
    bottom} panels are residual plots shown as functions of
  $L_{IR}^{CE}$ ({\it lower-left}) and redshift ({\it lower-right}).
  This figure shows that $L_{IR}^{CE}$, adopted as our best-fit
  bolometric luminosity is consistent with the range of $L_{IR}^{DH}$
  that one obtains if no $L_{MIR}-L_{IR}$ correlation is assumed.
\label{LIRvTIR}}
\end{figure*}

In Figure~\ref{LIRvTIR}({\it top}), $L_{IR}^{CE}$ ($L_{IR}$
based on CE01 templates) is plotted against the family of
$L_{IR}^{DH}$ values ({\it dots}).
The bottom panels are the residual plots shown as functions of the
$L_{IR}^{CE}$ ({\it lower-left}) and redshift ({\it lower-right}).
The mean ratio of unity for $L_{IR}^{DH}$/$L_{IR}^{CE}$ indicates that
the independently derived CE01 and DH02\nocite{Dale_Helou_02} $L_{IR}$
values are consistent on average.  The range of this ratio suggests a
mean LIR uncertainty of $\approx\!0.2$~dex for the sample as a whole,
and $\approx\!0.3$~dex for the most distant sources at $z\gtrsim1$.
At the $z\approx\!0.5$ the uncertainty is minimized, indicating that
the $L_{MIR}-L_{IR}$ correlation is tightest for galaxies measured at
restframe wavelengths of $\lambda\!\approx\!15\micron$.  It is worth
noting that $L_{IR}^{DH}$[3--1100\micron] and
$L_{IR}^{CE}$[8--1000\micron] are defined over different wavelength
ranges; however, since the flux between 3-8\micron\, and
1000-1100\micron\, is of order a few percent of the total bolometric
luminosity (DH02)\nocite{Dale_Helou_02} this difference is ignored
here.

\section{AGN Contamination}
\label{agn_sect}

The conversions of both emission line and $L_{IR}$ luminosity to a
star formation rate hinge on the assumption that the dominant
ionization and heating source is radiation from massive young stars.
AGN-dominated emission line and IR fluxes do not trace star formation
and must therefore be removed from our sample.  Various emission-line
diagnostics such as [OIII]/$H\beta$ versus [NII]/$H\alpha$ and
[OIII]/$H\beta$ versus [SII]/$H\alpha$ effectively separate
populations with different ionization sources \citep{Osterbrock_89}.
In cases with more limited spectral coverage, individual line ratios
such as [NeIII]/[OII] have also been successfully utilized
\citep{Kobulnicky_Kewley_04}.  We do not employ these line
diagnostics, due to the non-uniform rest-frame spectral coverage of
our sample.  Instead, we rely on an IRAC color selection and the
visual identification of optical spectral features to identify and
remove AGN candidates from our sample.

\begin{figure*}[!thp]
\epsscale{1.00}
\plotone{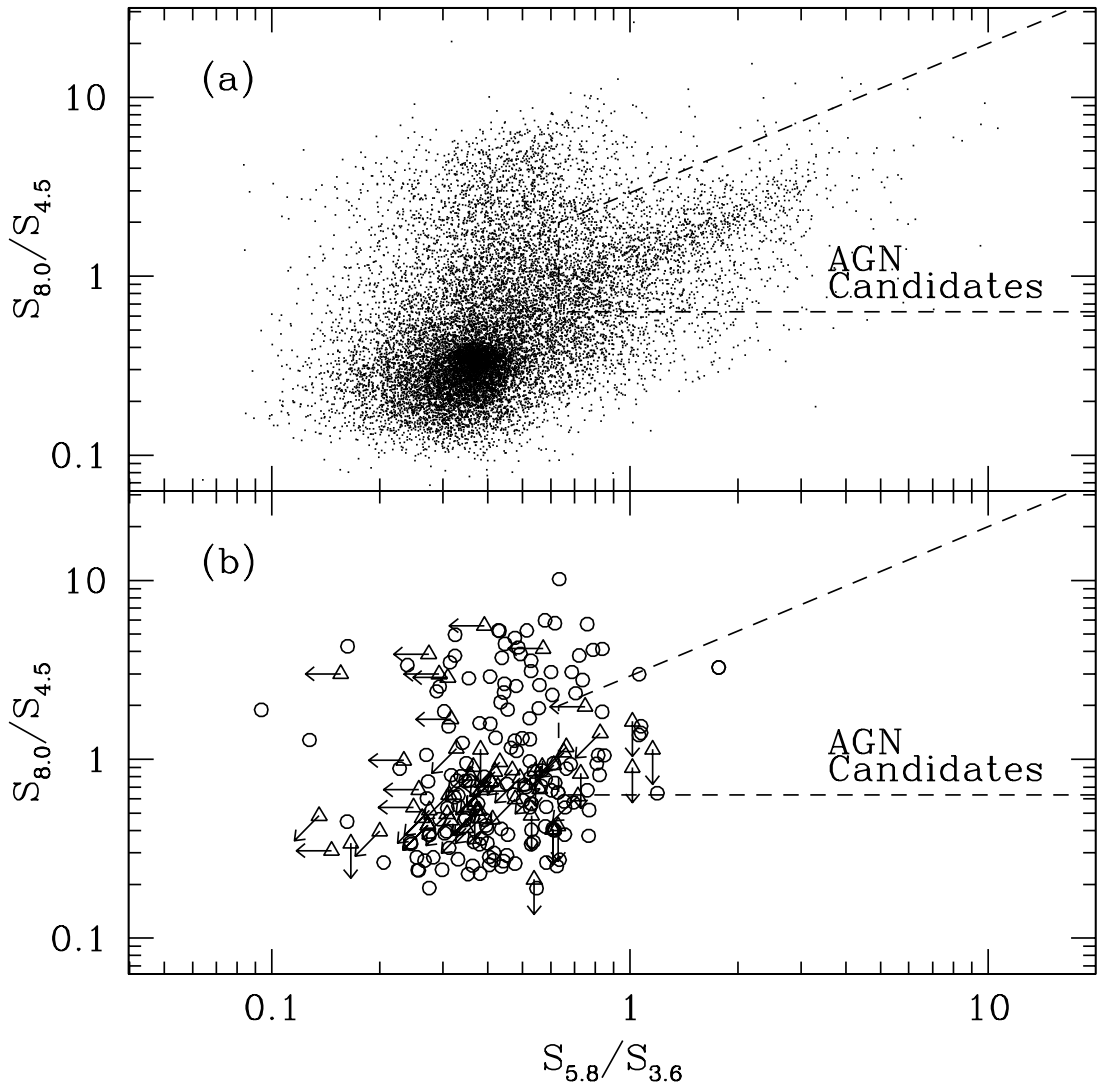}
\epsscale{1.0}
\caption{IRAC 4-channel color-color plots used to identify AGN
  candidates.  Panel~({\it a}) illustrates how the AGN candidate
  region of color-color space is defined by \citet{Lacy+04}.  All
  sources with clean 4-color photometry from the full 4~sq.~deg FLS
  main field are plotted; the region in which quasars and AGN are
  expected to be found is marked by a dashed line \citep*[cf.][for
  details]{Sajina+05}.  Panel~({\it b}) shows our sample with the same
  AGN region marked.  The majority of the sample ($\approx\!2/3$) has
  clean photometry in all four IRAC bands ({\it open circles}).  The
  remaining sources ($\approx\!1/3$) have at least two measured fluxes
  and therefore at least one limiting color ({\it open triangles}).
  Sources that fall in the dashed region of this color-color diagram
  are flagged as AGN candidates and are removed from our final sample.
\label{irac_cc_2panel}}
\end{figure*}

We first flag sources with obvious AGN signatures such as broadened
Balmer and/or [OII] lines or strong NeIII or NeV emission.  Next, we
combine our sample with IRAC photometry of the FLS \citep{Lacy+05} to
implement a 4-band IRAC color selection.  It has been shown that AGN
can be identified in the MIR, based on their strong continuum flux
\citep{Laurent+00, Lacy+04, Stern+05}.  \citet{Lacy+04} use 4-band
{\it Spitzer} IRAC photometry (3.6, 4.5, 5.8 \& 8.0\micron) to
identify a distinct region in color-color space where quasars and AGN
are likely to be found.  In Figure~\ref{irac_cc_2panel}{\it a}, we
reproduce the IRAC color-color plot from \cite{Lacy+04} (Figure 1) for
the entire FLS main field.  The dashed line shows the region expected
to be occupied by AGN.  In Figure~\ref{irac_cc_2panel}{\it b} we
show the same plot for our current sample.  The FLS depth of the IRAC
channels 3 \& 4 is slightly shallower than that of channels 1 \& 2,
and only 2/3 of our sample have clean photometry in all four IRAC
bands ({\it open circles}); for the remaining 1/3 of the sample,
limiting flux ratios are shown ({\it open triangles}).  Based on the
IRAC color selection, $9\%$ of each of these subsamples fall in the
AGN-candidate region.  Visual re-inspection of the 16 IRAC-selected
candidates, reveals that $63\%$ exhibit some AGN signature in their
spectra (6 strong; 4 moderate).  The remaining sources show no obvious
indicator; however, an AGN contribution cannot be ruled out, given our
limited spectral coverage.  We remove all 16 candidates from the final
sample.  Investigation of the spectroscopically selected candidates
shows that 5 of 9 sources (55\%) with strong AGN spectral signatures
are also IRAC-selected AGN.  Ultimately, we find that neither the
spectroscopic nor IRAC color selection provides a complete census of
all AGN, so we take the conservative approach of combining the two
methods to clean our sample.  The impact of this AGN removal is best
illustrated in the comparison of the two different SFR diagnostics
(Fig.~\ref{SFR2_noagn}), discussed in the next section.

\section{Optical vs. IR SFR comparison}
\label{opt_vs_IR_SFR}
\begin{figure*}[!thp]

\epsscale{1.00}
\plotone{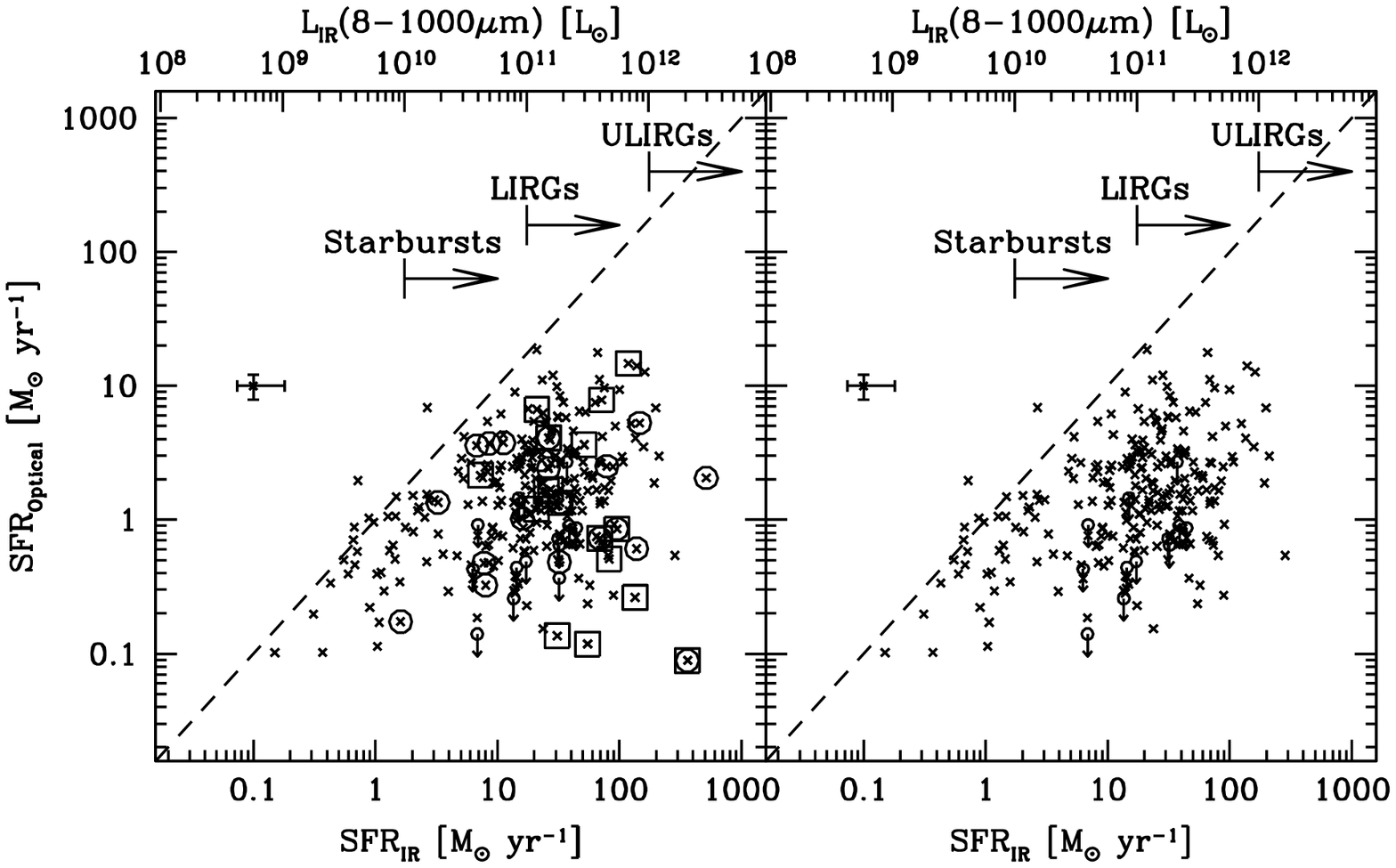}
\epsscale{1.0}
\caption{Comparison of $SFR_{opt}$ and $SFR_{IR}$ before ({\it left}) and after ({\it
    right}) AGN removal.  In the {\it left} panel a sample of 274
  sources with well-determined emission lines and IR luminosity are
  plotted as {\it crosses}.  The 33 AGN candidates identified by their
  spectral features ({\it squares}) and IRAC colors ({\it circles})
  are marked.  In the {\it right} panel, the final 241 source sample
  after AGN removal is shown.  No extinction correction has been
  applied to $SFR_{opt}$ in either panel.
  Representative error bars are shown in each panel.  The $SFR_{opt}$
  error bar is the mean $1\sigma$ uncertainty due to the error in line
  flux measurement.  The $SFR_{IR}$ error bar is the mean of the
  bolometric correction uncertainty, as discussed in
  \S\ref{sect_bol_corr}.  The comparison of these samples illustrates
  1) the importance of limiting the AGN contamination and 2) that
  without any reddening correction, $SFR_{opt}$ severely
  underestimates the true star formation rate, especially for our most
  IR luminous sources.
\label{SFR2_noagn}}
\end{figure*}

Next, we compare the two independently derived SFR diagnostics
$SFR_{opt}$ and $SFR_{IR}$ described in \S\ref{section_sfr} for the
full AGN-cleaned sample.  In Figure~\ref{SFR2_noagn}, we show the
reddening-uncorrected, emission line-derived $SFR_{opt}$ versus the IR
luminosity-derived $SFR_{IR}$, before and after AGN removal.  The
sample plotted in Figure~\ref{SFR2_noagn}~({\it left}) includes the
274 sources with well determined spectroscopic redshifts, emission
line and 24\micron\, fluxes.  AGN candidates are marked based on their
spectral ({\it squares}) or IRAC color identification ({\it circles})
(as discussed in \S\ref{agn_sect}).  In Figure~\ref{SFR2_noagn}~({\it
  right}) only the final 241~source AGN-removed sample is shown.
Though the distribution of the AGN candidates are not localized to a
single region of the plot, sources with the most extreme SFR
discrepancies tend to be AGN.  This illustrates the potential for
these sources to be misidentified as heavily obscured galaxies.  The
inclusion of this population would bias the sample to appear overly
obscured.  The fact that AGN candidates are found throughout the SFR
vs SFR plot is not unexpected since the emission line luminosity and 
IR luminosity are loosely correlated in AGN.

\begin{figure*}[!thp]
\epsscale{1.00}
\plotone{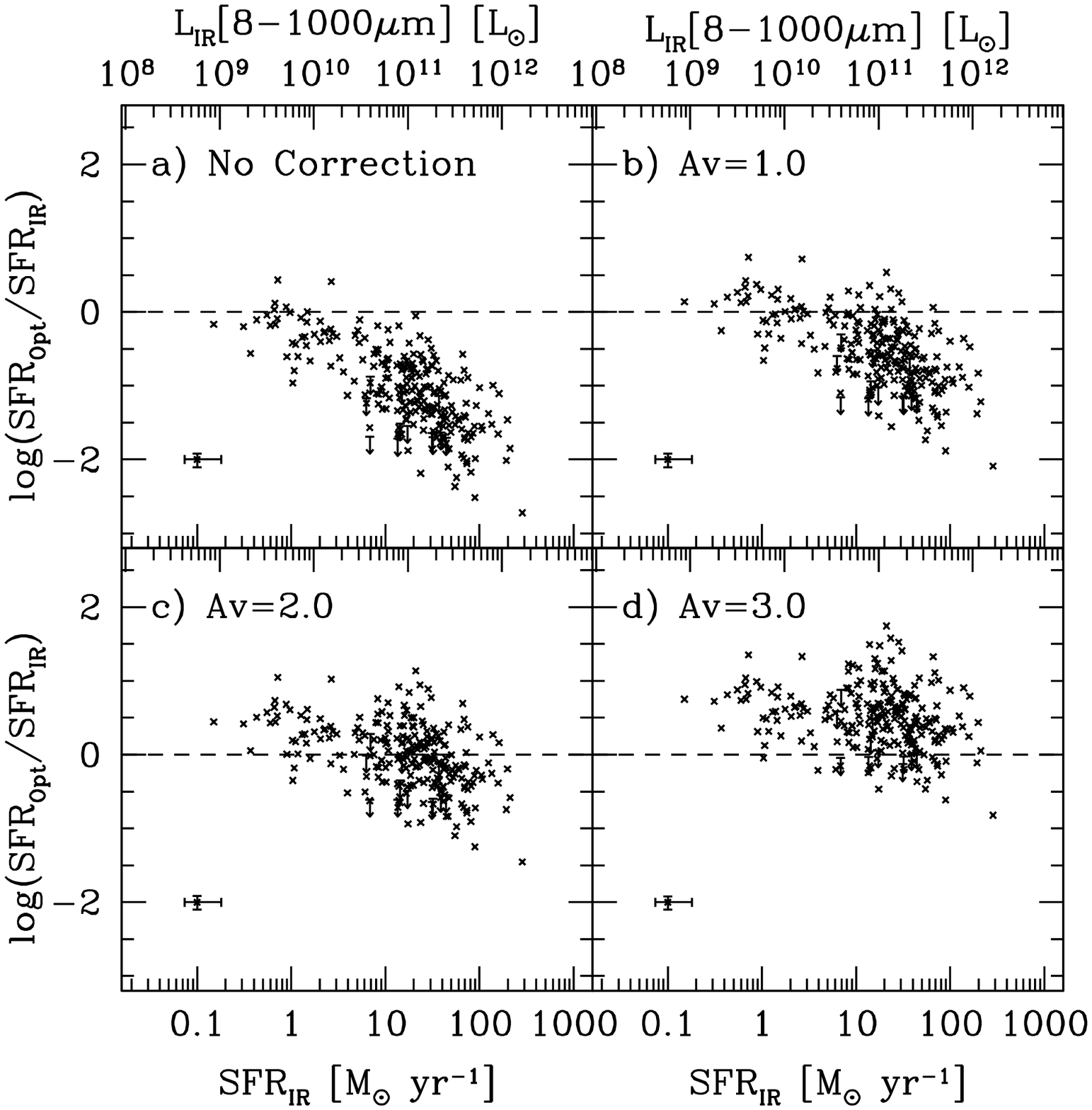}
\epsscale{1.0}
\caption{$SFR_{opt}/SFR_{IR}$ vs. $SFR_{IR}$ and $L_{IR}$ for the final
  AGN-cleaned sample, shown for a range of constant extinction corrections.
  In panel~({\it a}) no correction for any optical extinction is made.
  In panel~({\it b}) a uniform standard correction of $A_v=1.0$~mag,
  typical of normal spiral galaxies, is applied to $SFR_{opt}$.  In
  panels~({\it c \& d}) more extreme constant extinction corrections,
  $A_v=2.0$~mag and $A_v=3.0$~mag are adopted.  Representative error bars are
  shown in each panel.  The $SFR_{opt}/SFR_{IR}$ error bar includes
  only the mean $1\sigma$ uncertainty due to the error in line flux
  measurement, it does not include any uncertainty in $SFR_{IR}$.
  The $SFR_{IR}$ error bar is the mean of the bolometric correction
  uncertainty, as discussed in \S\ref{sect_bol_corr}.  With no
  extinction correction, $SFR_{opt}$ underestimates the true star
  formation rate by as much as 2.5~dex for our most IR luminous
  sources.  Uniform extinction corrections do not adequately represent
  our sample of normal, starburst and LIRG galaxies.
\label{DELTA_SFR2_4panel_constAv}}
\end{figure*}

In Figure~\ref{SFR2_noagn}~({\it right}) the uncorrected $SFR_{opt}$
is systematically lower than $SFR_{IR}$ .  Over the $L_{IR}$ range of
our sample, the offset spans 2 orders of magnitude and illustrates the
importance of properly accounting for the optical extinction.  In this
section we explore a range of different extinction corrections to
reconcile the two SFRs.  In Figure~\ref{DELTA_SFR2_4panel_constAv},
the ratio of the two diagnostics, $SFR_{opt}$/$SFR_{IR}$ is plotted to
better illustrate the systematic difference between them.  It is
evident from Figure~\ref{DELTA_SFR2_4panel_constAv}{\it a}, in which
no reddening correction has been applied that $SFR_{opt}$
underestimates the true SFR by anywhere from $0-2.5$~dex.  In the
remaining panels, we investigate the applicability of various fixed
extinction corrections.  In
Figure~\ref{DELTA_SFR2_4panel_constAv}({\it b}), an extinction of
$A_v=1.0$, representative of normal spiral galaxies is assumed.  This
is a significant improvement over the uncorrected $SFR_{opt}$,
especially at low IR luminosity, typical of normal late-type galaxies.
Beyond $L_{IR}>10^{10} L_{\odot}$, however, the two $SFRs$ diverge,
indicating that this standard correction underestimates the
attenuation in starburst and brighter galaxies.  In
Figures~\ref{DELTA_SFR2_4panel_constAv}{\it c \& d}, we adopt more
extreme corrections of $A_v=2.0$~mag and $A_v=3.0$~mag and find that neither
provides a good solution over the full range in $L_{IR}$.  The
\citet{Calzetti+00} reddening curve is adopted for each of these
corrections.  We conclude that a constant $A_v$ correction is
insufficient for reconciling $SFR_{opt}$ and $SFR_{IR}$.  In the
following sections, we take two independent approaches to characterize
the optical extinction of galaxies of our sample.  In the first, we
use Balmer decrement and other emission line ratios to compute
extinction, $A_v^{em}$.  In the second, we adopt $SFR_{IR}$ as a
proxy for the true total star formation rate and compute $A_v^{IR}$
based on the difference between $SFR_{opt}$ and $SFR_{IR}$.

\subsection{Optical Extinction Correction I: Balmer Decrement}

A standard method for measuring the internal extinction of an
individual galaxy is through the comparison of observed to predicted
Balmer line ratios.  We adopt this approach for the subset of our
sample with multiple measured Balmer lines.  The color excess
$E(B-V)_{gas}$ of a source is computed by comparing the observed
Balmer line ratios
$(F_o^{H\alpha}/F_o^{H\beta})$,
$(F_o^{H\delta}/F_o^{H\beta})$ and 
$(F_o^{H\gamma}/F_o^{H\beta})$ with their intrinsic unobscured ratios:

\begin{equation}
E(B-V)_{gas}=\frac{2.5}{[k(\lambda_1)-k(\lambda_2)]}\log\frac{(F_i^{\lambda_1}/F_i^{\lambda_2})}{(F_o^{\lambda_1}/F_o^{\lambda_2})}
\end{equation}

\noindent where ($F_i^{\lambda_1}/F_i^{\lambda_2}$) are the intrinsic
unobscured line ratios
based on case-B recombination and $T=10,000$~K
\citep{Osterbrock_89}.  The obscuration or reddening curve
($k(\lambda)$) has been derived by \cite{Calzetti+00} for starburst
galaxies:

\begin{equation}
  k(\lambda)=
  \cases{
    2.659(-1.857+1.040/\lambda)+R_v\cr
    \qquad (0.63\micron\leq\lambda\leq2.20\micron) \cr
\cr
    2.659(-2.156+1.509/\lambda\cr
    -0.198/\lambda^2+0.011/\lambda^3)+R_v\cr
    \qquad (0.12\micron\leq\lambda\leq0.63\micron)
  }
\end{equation}

\noindent It should be noted that although there are significant
large-scale differences between the various starburst, Milky Way, LMC
and SMC reddening curves, over the rest-frame spectral region covered
by the emission lines of interest
(3700\AA$\lesssim\lambda\lesssim$6600\AA), the differences are minor
\citep[][and references therein]{Calzetti_01}.  The color excess is
converted into a wavelength dependent extinction based on:

\begin{equation}
  A(\lambda)=E(B-V)_{gas}k(\lambda)
\end{equation}

\noindent where A($\lambda$) is the mean emission line derived
extinction in units of magnitudes at the wavelength $\lambda$.  This
should be distinguished from $A_s(\lambda)$, the extinction as
measured from the stellar continuum.  As has been shown by
\citet{Calzetti_01} these differ by a factor of 0.44 in typical local
starbursts.

\begin{figure*}[!thp]
\epsscale{1.00}
\plotone{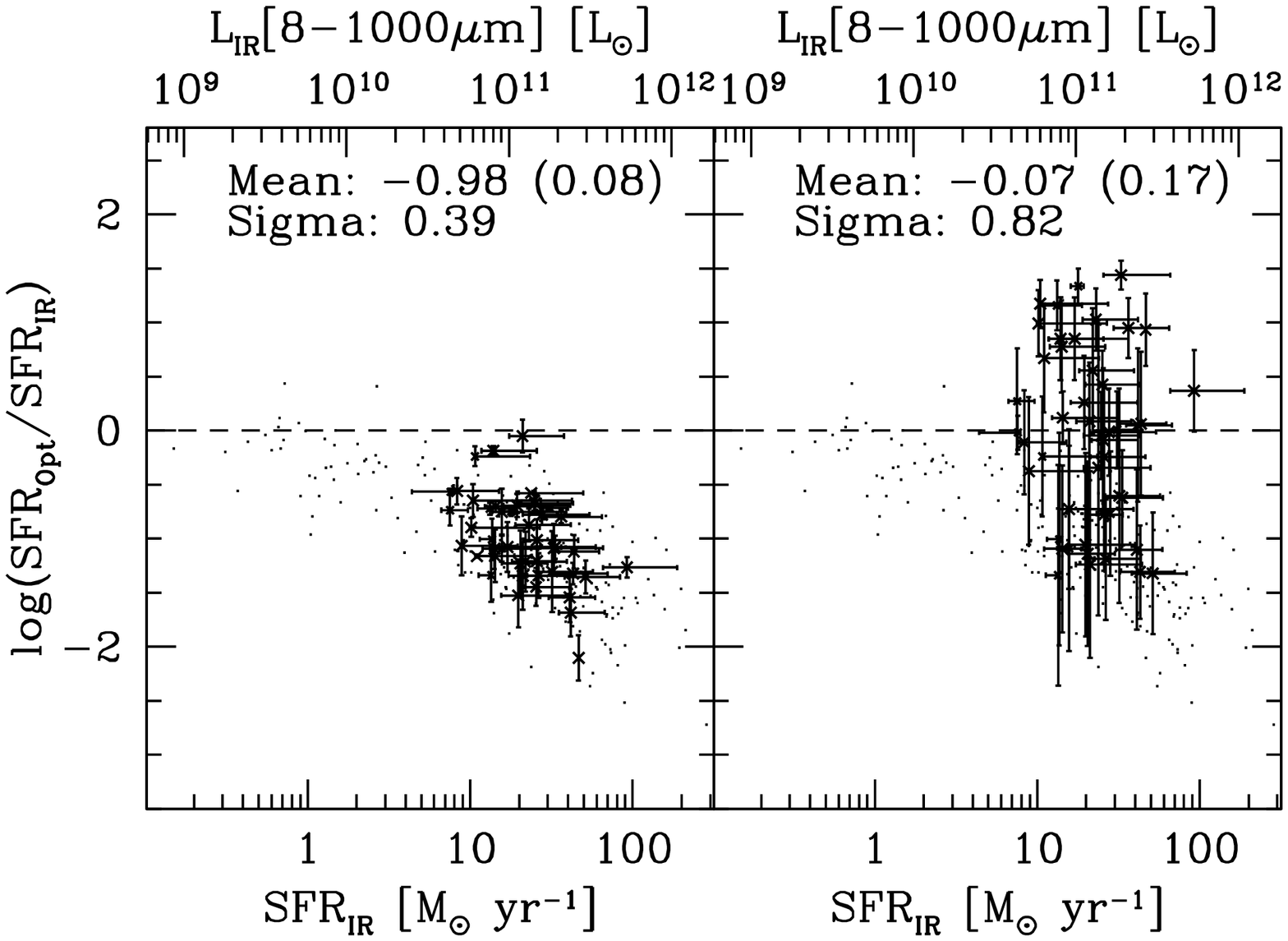}
\epsscale{1.0}

\caption{$SFR_{opt}/SFR_{IR}$ versus $SFR_{IR}$ and $L_{IR}$ before
  and after application of {\it emission line ratio}-derived reddening
  corrections, $A_v^{em}$.  A sample of 24 sources is shown ({\it
    crosses}) before ({\it left}) and after ({\it right}) the
  application of $A_v^{em}$.  For reference, the full sample is also
  plotted ({\it small dots}).  The $SFR_{IR}$ error bars represent the
  full range of the bolometric correction uncertainty, as discussed in
  \S\ref{sect_bol_corr_unc}.  The $SFR_{opt}/SFR_{IR}$ error bars
  show the $1\sigma$ uncertainty due to the error in line flux
  measurement, excluding any contribution from $SFR_{IR}$.  In
  panel~{\it b}, individual y-error bars also include the uncertainty
  in $A_v^{em}$.   Application of $A_v^{em}$ results in a reasonable 
{\it ensemble-averaged} SFR, as evidenced by the improvement of $\langle log(SFR_{opt}/SFR_{IR})\rangle$ from -0.98 to -0.07.   However, the significant increase in scatter suggests that our emission line ratios do not provide optimal corrections for individual sources.
\label{DELTA2_bdec_2panel}}
\end{figure*}

In addition to using the Balmer decrement, we explore the possibility
of using the [OII]/$H\beta$ line ratio to compute $A_v^{em}$.  Though
this ratio is less certain than that of the Balmer lines, we adopt an
empirical value for the intrinsic [OII]/$H\beta$ line ratio based on the
locally measured reddening-corrected line fluxes.  Using the NFGS,
\citet*{Kewley+04} measure a mean extinction-corrected [OII]/$H\alpha$
ratio of 1.2, which translates to an [OII]/$H\beta$ line ratio of:

\begin{eqnarray*}
  F_i^{[OII]}/F_i^{H\beta}&=3.44\\
\end{eqnarray*}

\noindent From our full sample, individual reddening corrections,
$A_v^{em}$, are computed for 24 sources that have multiple strong
Balmer and/or [OII] emission lines.  Line fluxes and $SFR_{opt}$ are
then reddening-corrected on a source-by-source basis.  In
Figure~\ref{DELTA2_bdec_2panel}, $SFR_{opt}/SFR_{IR}$ vs.  $SFR_{IR}$
for this sample ({\it crosses}) is shown before ({\it left}) and after
({\it right}) application of this correction.  
There is a significant increase in the scatter of the distribution
after application of the extinction correction.  This illustrates that
due to the enormous uncertainties associated with $A_v^{em}$,
source-by-source corrections are futile.  Despite the increased
scatter, it is noteworthy that $SFR_{opt}$ becomes consistent on
average with $SFR_{IR}$, with the mean offset reduced from -0.98~dex
to -0.07~dex.  This indicates that in the absence of other extinction
indicators, with a large enough sample and careful stellar absorption
measurements, Balmer decrement and [OII]/$H\beta$ emission line ratios
can provide a reasonable first order $\langle A_v^{em} \rangle$
correction for the luminosity range of this sample.  The median
emission line derived extinction of this subsample is
$\tilde{A}_v^{em}=1.5\pm1.1$~mag with $\langle
SFR_{IR}\rangle=23\pm8~M_{\odot}yr^{-1}$.  Due to the relatively small
size and limited $L_{IR}$ coverage of this subsample, along with the
large uncertainties, we are not able to derive an $L_{IR}$-dependent
extinction correction.  Instead, in the next section we adopt
$SFR_{IR}$ as a `true' SFR and compare it to the
extinction-uncorrected $SFR_{opt}$ to derive $A_v^{IR}$ for the full
sample.

\subsection{Optical Extinction Correction II: IR vs Optical SFRs}
\subsubsection{Computing $A_v^{IR}$}

In this section, we take an alternative approach of adopting
$SFR_{IR}$ as a proxy for the true SFR to determine the extinction
correction of our sample.  Specifically, the ratio $SFR_{IR}$ over the
uncorrected $SFR_{opt}$ is used to compute the attenuation of the
given emission line:

\begin{equation}
A(\lambda)=-2.5*\log(SFR_{opt}/SFR_{IR})
\end{equation}

\noindent where $\lambda$ is the wavelength of the emission line used
to compute $SFR_{opt}$.  $A(\lambda)$ is converted to a standard
visual extinction, $A_v$ by:

\begin{equation}
A_v=\frac{A(\lambda)}{k(\lambda)}k_v
\end{equation}

\begin{figure*}[!h]
\epsscale{1.00}
\plotone{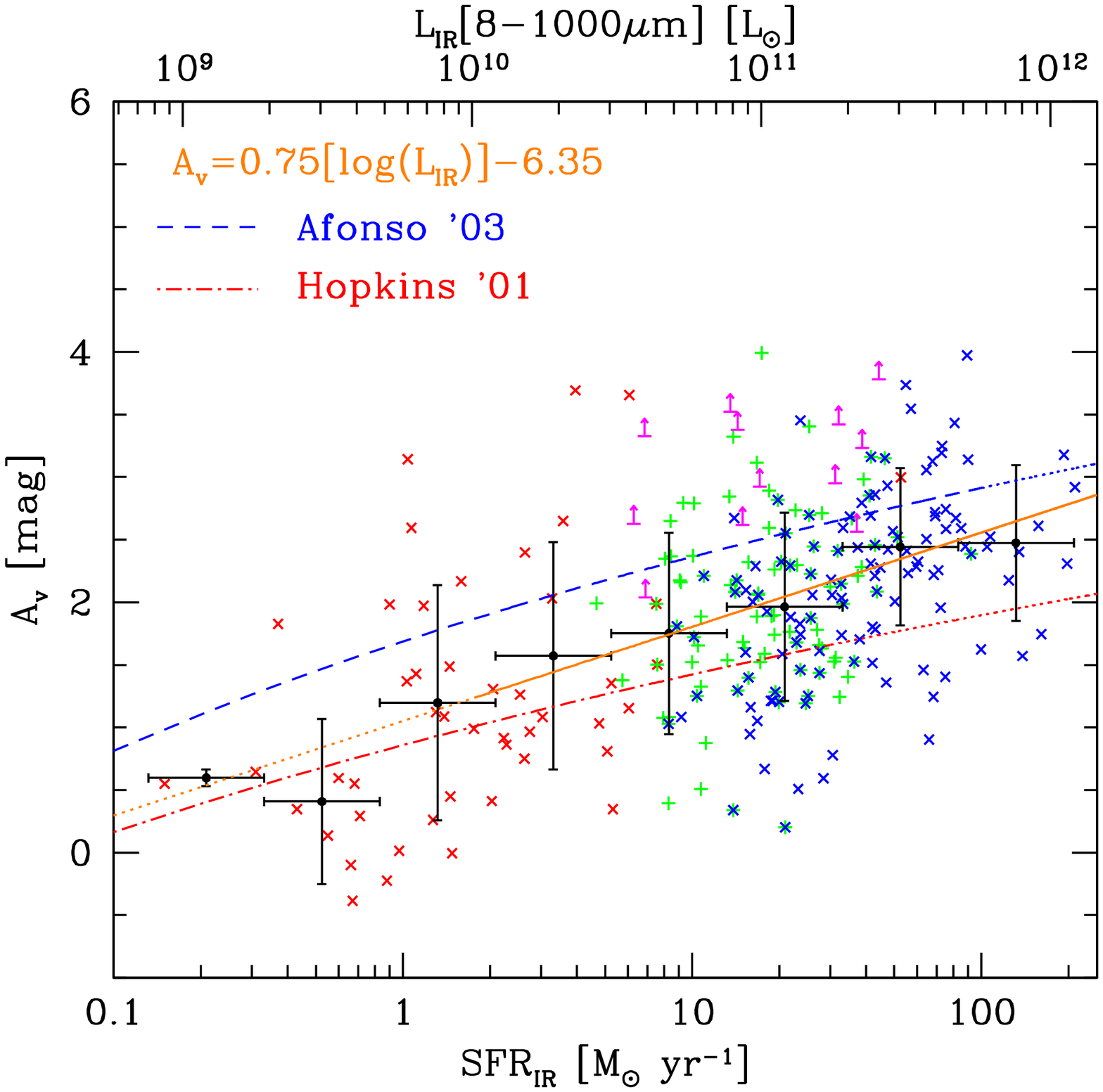}
\epsscale{1.00}
\caption{Optical extinction $A_v^{IR}$ vs. $L_{IR}$ for the complete
  galaxy sample.  $A_v^{IR}$ is derived from the ratio of
  $SFR_{opt}/SFR_{IR}$, where $SFR_{IR}$ is adopted as a proxy for the
  true SFR, and plotted as a function of IR luminosity ({\it crosses
    \& pluses}).  Sources with measured $H\alpha$, $H\beta$ \& [OII]
  line fluxes are shown as {\it red}, {\it green} and {\it blue}
  points, respectively.  Those with only emission line flux limits are
  shown as {\it magenta arrows}.  Mean $A_v^{IR}$ values ({\it solid
    circles}), binned in $log(SFR_{IR})$ are plotted along with their
  $1\sigma$ uncertainties.  The best-fit line
  $A_v^{IR}=0.75*log(L_{IR}/L_{\odot})-6.35$~mag is computed and overlaid
  as a {\it solid orange} line and compared to those derived by
  \citet{Hopkins+01} ({\it dot-dashed red}) and \citet{Afonso+03}
  ({\it dashed blue}).  Extrapolations beyond the dataset luminosity
  limits are indicated by {\it dotted} lines.  The mean measured
  $A_v^{IR}$ function of our sample is intermediate between those of
  H01 \& A03, with evidence for a stronger $L_{IR}$ dependence.
\label{Av_function_ranga}}
\end{figure*}

\noindent Extinction derived in this manner may have a large 
uncertainty for individual sources, but this approach should produce
a reasonable ensemble average.  In Figure~\ref{Av_function_ranga},
$A_v^{IR}$ versus $SFR_{IR}$ and $L_{IR}$ is shown for our full
sample.  Despite the large scatter in the distribution, a clear trend
of increasing $A_v^{IR}$ as a function of $SFR_{IR}$ is evident.  The
best-fit line to this distribution is:

\begin{equation}
  A_v^{IR}=0.75(\log(SFR_{IR})) + 1.05
\end{equation}

\noindent which in terms of $L_{IR}$ is:
\begin{equation}
A_v^{IR}=0.75*\log(L_{IR}/L_{\odot})-6.35
\end{equation}

\noindent This fit is limited to starburst and brighter systems 
($L_{IR}>10^{10} L_{\odot}$; {\it orange solid line}), where
$SFR_{IR}$ is a good representation of the total SFR.  It is
extrapolated to lower luminosities ($L_{IR}<10^{10} L_{\odot}$; {\it
  orange dotted line}).  Inclusion of the full sample has only a minor
effect, slightly steepening the fit.  The mean $A_v^{IR}$ values ({\it
  black dots}) of sources binned in $log(SFR_{IR})$ are overlaid along
with error bars that show their $1\sigma$ dispersion.

\subsubsection{Comparison to Local Samples}
We make a direct comparison of our $A_v^{IR}$ function to those of
local optical/UV \citet{Hopkins+01} and radio-selected samples of
\citet{Afonso+03} (hereafter H01 \& A03).  We transform their E(B-V)
color excess to $A_v$ based on $R_v=3.1$ and plot them as {\it red
  dot-dashed \& blue dashed} curves, with dotted segments representing
extrapolations beyond their datasets.  For a given SFR, the mean
$A_v^{IR}$ of our sample is intermediate between those of H01 and A03.
The trend in the extinction as a function of $L_{IR}$ is also slightly
steeper than the previous relationships.  To interpret these
differences requires a closer examination of the various sample
selections.

The H01 SFR dependent extinction relationship is based on small
($N\!\approx\!60$) local UV+optical selected samples (cf.
\citet{Sullivan+01}), whereas that of A03 is derived from a comparably
sized radio-selected sample that extends to slightly higher redshift
$\langle z \rangle \approx\!0.25$ and is more actively starforming.
The difference in the mean $A_v$ between the two samples is attributed
to the fact that optical/UV selected samples are biased against
heavily obscured galaxies.  Our initial spectroscopic sample being
NIR-selected ($2.2\micron$), we expect our sample to be significantly
less affected by obscuration effects than that of a UV/optical sample.
It is therefore not surprising that our mean $A_v^{IR}$ is higher than
that of H01.

The discrepancy between our sample and that of A03 is more difficult
to understand since both radio and NIR/MIR selections should be
relatively robust against obscuration biases.  It is suggestive of a
selection bias in one or both samples.  Reliance on optical
spectroscopy may be one potential culprit.  It is possible that the
dustiest systems are so heavily obscured that either their redshifts
are indeterminable or their emission lines are completely attenuated.
We can place some constraints on the sizes of these two populations in
our sample.  Our spectroscopic redshift efficiency of $92\pm5\%$
places an 8\% limit on the first.  This is a conservative estimate
since some fraction of the 8\% are certainly due to our spectroscopic
redshift sensitivity function dropping off beyond $z\!\approx\!1.3$.
Regarding the second population, 12 sources in our sample ($<4\%$)
with identified redshifts, have no measured emission line fluxes and
therefore only lower limits on the derived attenuation.  These are
shown as arrows in Figures~\ref{SFR2_noagn} \& \ref{Av_function_ranga}
and included in our $A_v^{IR}(L_{IR})$ estimate.  Deeper optical
spectroscopy may reveal these to be even more heavily obscured
systems.  Based on these estimates, these two populations are likely
to account for $\lesssim10\%$ of the total population.  It is worth
noting that these biases are not unique to this study since most
surveys rely on optical spectroscopy for redshift and/or emission line
diagnostics.

One potential bias that may {\it overestimate} the mean optical
attenuation is due to AGN contamination.  Especially in the most
luminous samples, the contribution from AGN becomes increasingly
significant.  Radio-selected, and to a lesser extent, MIR-selected
samples will tend to have larger fractions of `active' galaxies than
comparable depth optical or NIR samples.  As seen in
Figure~\ref{SFR2_noagn}, if this population is not sufficiently
screened, it can result in an overestimate of the sample $A_v$.

\begin{figure*}[!thp]
\epsscale{1.00}
\plotone{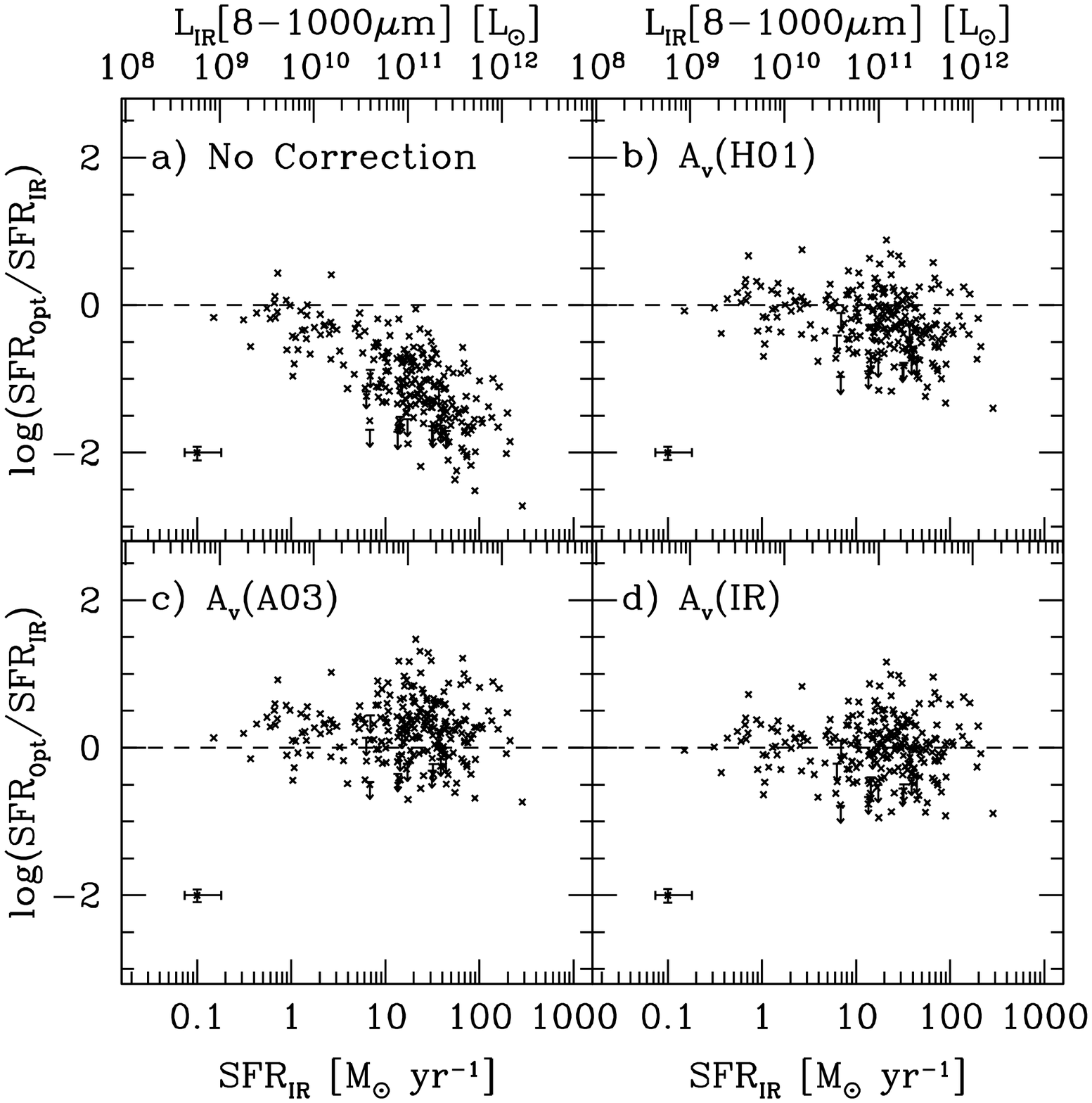}
\epsscale{1.00}
\caption{$SFR_{opt}/SFR_{IR}$ vs. $SFR_{IR}$ and $L_{IR}$ of the final
  AGN-cleaned sample, with various luminosity dependent extinction
  corrections.  In panel~({\it a}) no correction for any optical
  extinction is made.  In panels~({\it b \& c}) we adopted the $A_v$
  functions derived by H01 and A03 respectively.  In panel~({\it d})
  we adopt our best-fit $A_v^{IR}$ extinction function based on our
  sample.  Representative error bars are the same as in
  Figure~\ref{DELTA_SFR2_4panel_constAv}.  We see that the corrections
  of H01 and A03 appear to under/over-estimate the mean extinction of
  our sample.  The scatter of the corrected sample in panel~({\it d})
  is 0.3~dex.
\label{DELTA_SFR2_4panel_best}}
\end{figure*}

Finally, our $A_v^{IR}$ values appear to be consistent with those
derived from an SDSS sample of starforming galaxies
\citep{Hopkins+03}.  Though no direct comparison is shown here,
inspection of their Figure~8 shows that the mean $A_v$ of their sample
is intermediate between that of H01 and A03, consistent with the
correction derived here over a comparable range of SFR.

\subsubsection{Disentangling the $L_{IR}$ vs. Redshift Dependence}
\label{redshift_dependence}
As seen in Figure~\ref{LIR_vs_z}, our sample spans a broad range in
both IR luminosity and redshift; and as a consequence of our 24\micron\,
flux limit it exhibits a strong $L_{IR}$-redshift correlation.  So
far, we have been assuming that for a fixed $L_{IR}$, $A_V$
is largely redshift independent.  Given the potential degeneracy
between redshift and IR luminosity dependencies, the validity of this
assumption merits investigation.

\begin{figure*}[!thp]
\epsscale{1.00}
\plotone{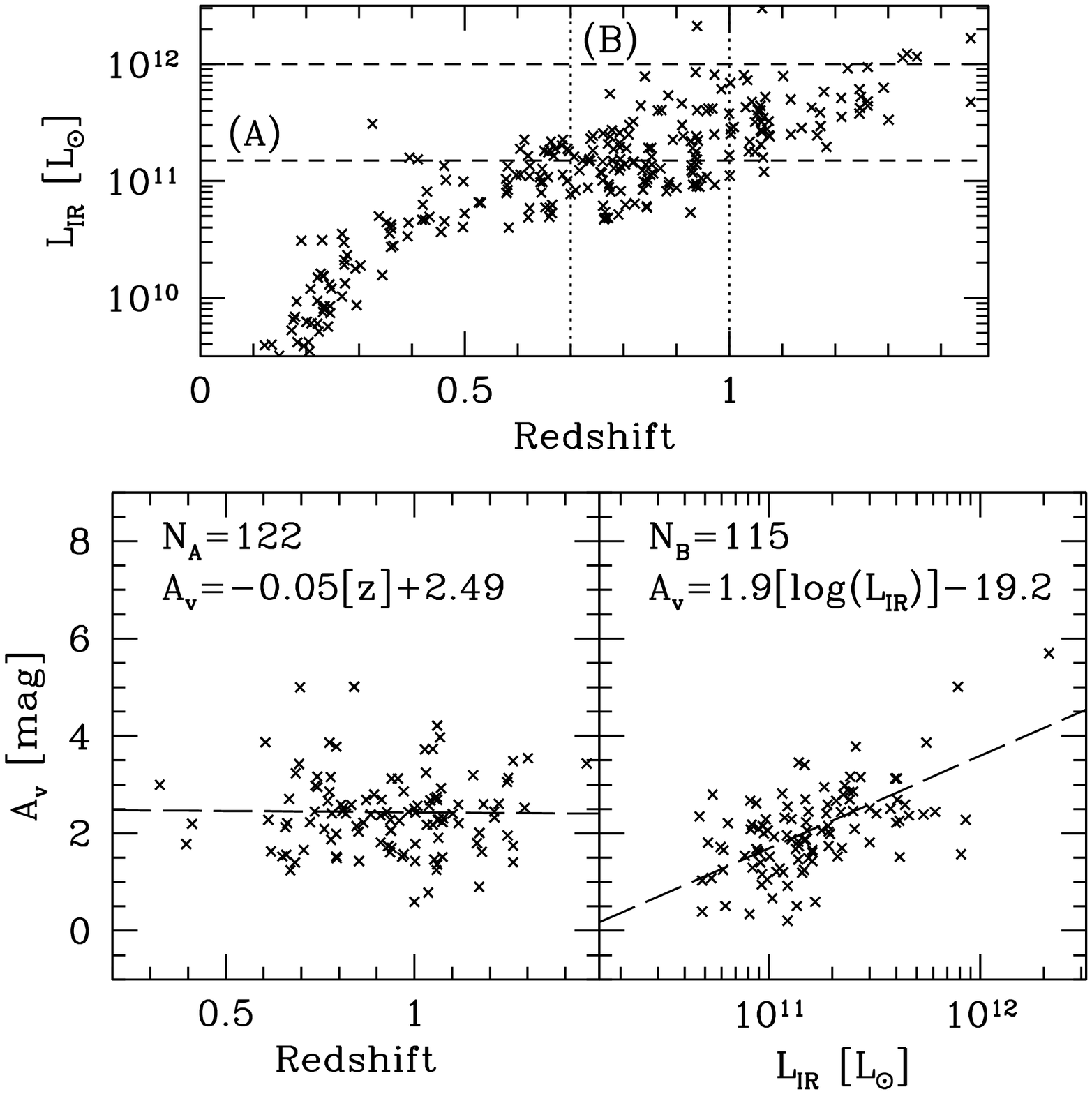}
\epsscale{1.00}
\caption{A comparison of the $A_V^{IR}$ dependence on IR luminosity and redshift for two subsets
of our final sample.  In the $L_{IR}$ versus redshift plot ({\it top}), two regions limited in 
$L_{IR}$ ($2\times10^{11}<L_{IR}<1\times10^{12}$) and redshift ($0.7<z<1.0$) are identified 
with {\it dashed \& dotted} lines and marked {\it A \& B}, respectively.  The lower 
panels show $A_V^{IR}$ vs. redshift for 122 galaxies from region {\it A} 
({\it lower-left}) and $A_V^{IR}$ vs. $L_{IR}$ for 115 galaxies from
region {\it B} ({\it lower-right}).
The {\it lower-left} panel reveals that for a fixed $L_{IR}$, $A_V$ exhibits almost 
no redshift dependence.  By contrast, the {\it lower-right} panel shows that even 
over a narrow redshift range, $A_V$ has a clear $L_{IR}$ dependence.
\label{Av_vs_zLIR_slices}}
\end{figure*}

Though our sample is not ideally suited for a thorough study of
this issue, we can place some limits on the redshift
dependence by isolating subsets of our data restricted in IR
luminosity and redshift.  In Figure~\ref{Av_vs_zLIR_slices}, we take
a look at two such slices in $L_{IR}$ and redshift.  The first is
restricted to $2\times10^{11}<L_{IR}/L_{\odot}<1\times10^{12}$, and
the second to $0.7<z<1.0$.  In the lower panels, plots of $A_V$ vs.
redshift and  $A_V$ vs. $L_{IR}$ for the respective subsets reveal that a)
for a sample with a narrow range of $L_{IR}$, there is no strong trend
in $A_V$ with redshift; and b) over a relatively narrow redshift slice
centered at $z=0.85$, $A_V$ shows a clear correlation with $L_{IR}$.
Though our sample does not allow us to investigate trends over the
entire redshift or $L_{IR}$ range of our sample, these two slices
suggest that to first order the dependence on redshift is small
compared to that on $L_{IR}$.

\section{Summary}
We have combined MIR (24\micron) photometry with high-resolution,
optical spectroscopy for a large ($N\,=\,241$), $Ks+MIPS\,
24\micron$-selected ($K_s\,<\,20.2$~mag; $f24\,>\,80\mu Jy$) galaxy
sample.  AGN are removed by implementing both spectroscopic and MIR
color selections.  This dataset is used to measure the instantaneous
star formation rate and the mean attenuation of normal through IR
luminous galaxies ($10^9<L_{IR}<10^{12.5} L_{\odot}$; $\langle L_{IR}
\rangle=1.8\times10^{11} L_{\odot}$; $\langle SFR \rangle=31~M_{\odot}
yr^{-1}$ ) out to a redshift of $z<1.5$ ($\langle z \rangle=0.77$).  We
compare two independent approaches of computing the star formation
rate.  The first is based on the IR luminosity, $SFR_{IR}$, and the
second is based on optical Balmer and [OII] emission line fluxes,
$SFR_{opt}$.

Comparison of the two SFR diagnostics reveal that with no correction
for extinction, the optical SFR systematically underestimates the IR
SFR by as much as 2.5~dex.  This discrepancy has a clear IR luminosity
dependence that cannot by reconciled with a constant $A_v$ extinction
correction.  We take two independent approaches to investigate the
dust attenuation of our sample.

First, we compute Balmer decrement and emission line ratio derived
optical extinction $A_v^{em}$ on a source-by-source basis for a subset
of our sample.  We find that after correction, despite a large scatter
in the distribution, the ensemble averaged $SFR_{opt}$ is consistent
with $SFR_{IR}$.  The large errors associated with the derived
$A_v^{em}$, however, illustrate that even with the stellar continuum
properly measured, the high-order Balmer line ratios, such as
$H\beta/H\gamma$ and $H\beta/H\delta$, as well as [OII]/$H\beta$ have
only a limited usefulness for extinction corrections of individual
galaxies at high redshift.  This is due to: a) the relative weakness
of these high order emission lines; and b) the limited leverage
afforded by the narrow separation of these lines, in comparison to the
conventionally adopted $H\alpha$/$H\beta$ Balmer decrement.

As an alternative measure of the optical attenuation, we use
$SFR_{opt}/SFR_{IR}$, to derive an IR-luminosity dependent extinction
function, $A_v^{IR}=0.75*log(L_{IR}/L_{\odot})-6.35$~mag.  In comparing
this relationship with local optical/UV and radio-selected samples, we
find that our $A_v^{IR}$ function is intermediate between those of
\citet{Hopkins+01}, \citet{Sullivan+01} and \citet{Afonso+03} but with
a slightly steeper slope.  Though it is difficult to reconcile these
results without a directly overlapping sample, we highlight some of
the key differences that may contribute to the discrepancy.
\citet{Afonso+03} suggests that the difference between their results
is due to selection bias against highly obscured objects in the
original \citet{Hopkins+01} sample.  Our sample should be relatively
free of this selection bias and therefore comparable to that of
\citet{Afonso+03}.  The fact that they are not is a bit surprising,
but may suggest that one or both of these samples is still suffering
from an additional bias.  One possibility is that we may be excluding
the most highly obscured sources due to our spectroscopic
sensitivities.  This is not likely to have a dominant effect given the
high $92\%$ spectroscopic redshift efficiency of our sample and the
fact that all spectroscopic samples are affected by this bias.
Another possibility is that the \citet{Afonso+03} sample, due to its
radio-selected nature may contain a larger fraction of galaxies with
AGN that may be inflating their mean $A_v$.  The inclusion of even
non-dominant AGN may have a similar effect on the measured ensemble
extinction as that of a heavily dust-obscured population.

We do not attempt to quantify the $A_V$ evolution over our full
redshift range.  However, based on subsamples restricted to narrow
redshift and $L_{IR}$ ranges, we find no evidence for significant
$A_V$ evolution.

The scatter in $SFR_{opt}/SFR_{IR}$ after application of our best-fit
attenuation correction, is larger than expected from either the line
flux errors or the reasonable estimates of the IR bolometric
correction uncertainty.  The impact of adopting a mean metallicity
correction ($\lesssim0.1$~dex) also cannot account for this scatter.
This indicates that these diagnostics are not providing consistent
tracers of the current SFR, possibly due to differences in their
dependence on dust geometry and/or SFR timescale.

\section{Acknowledgements}

We thank the anonymous referee for insightful comments that have
significantly improved this paper.  We would also like to thank Carol
Lonsdale, Robert Kennicutt and Rolf Jansen for early discussions that
helped shape this project.  We thank all of our colleagues associated
with the {\it Spitzer} mission, which has been carried out at the Jet
Propulsion Laboratory, operated by California Institute of Technology,
under NASA contract 1407.
We are indebted to Grant Hill, Greg Wirth and the rest of the Keck
observatory staff for their phenomenal observation support.  The
analysis pipeline used to reduce the DEIMOS data was developed by the
DEEP2 group at UC Berkeley with support from NSF grant AST-0071048.
Finally, we wish to recognize and acknowledge the significant cultural
role and reverence that the summit of Mauna Kea has within the
indigenous Hawaiian community.  We are grateful for the opportunity to
conduct observations from this mountain.

\bibliography{ms.bbl}


\end{document}